\documentclass[11pt]{article}
\pdfoutput=1

\usepackage{amsmath,amssymb,epsfig,amsfonts}
\usepackage{graphicx,subfigure}
\usepackage{epsf}
\usepackage[backref]{hyperref}
\usepackage[all]{xy}
\usepackage{slashed}
\usepackage{verbatim} 
\usepackage{float}
\restylefloat{table}
\usepackage[all]{xy}
\usepackage{pdflscape}
\usepackage{tikz}
\usepackage{array}
\usepackage{youngtab}
\usepackage{multirow}
\usepackage{color}
\usepackage{ulem}
\usepackage{cleveref}

\usepackage{tensor}

\makeatletter

\newcommand{\be}{\begin{equation}}
\newcommand{\ee}{\end{equation}}
\newcommand{\ba}{\begin{aligned}}
\newcommand{\ea}{\end{aligned}}

\newcommand{\Spin}{\mathrm{Spin}}
\newcommand{\CY}{\hbox{CY}}
\newcommand{\N}{\mathcal{N}}

\def\im{\mathop{\mathrm{im}}\nolimits}

\newcommand\Q{\mathbb{Q}}

\newlength{\sswidth}

\newcommand{\C}{\mathbb{C}}
\renewcommand{\P}{\mathbb{P}}

\newcommand{\bea}{\begin{eqnarray}}
\newcommand{\eea}{\end{eqnarray}}

\newcommand{\R}{{\mathbb R}}
\newcommand{\Z}{{\mathbb Z}}

\def\Im{\mathop{\mathrm{Im}}\nolimits}

\def\Re{\mathop{\mathrm{Re}}\nolimits}

\def\Tr{\mathop{\mathrm{Tr}}\nolimits}

\def\ker{\mathop{\mathrm{ker}}\nolimits}
\def\coker{\mathop{\mathrm{coker}}\nolimits}

\def\unit{{1\kern-.65ex {\rm l}}}
\def\1{{1\kern-.65ex {\rm l}}}

\def\bbE{{\mathbb{E}}}

\def\bbS{{\mathbb{S}}}

\newcount\hour \newcount\minute
\hour=\time \divide \hour by 60
\minute=\time
\def\now{%
\ifnum \hour<13
  \ifnum \hour=0 \advance \hour by 12 \number\hour:\else \number\hour:\fi%
     \ifnum \minute<10 0\fi%
     \number\minute%
\ A.M.%
\else \advance \hour by -12 \number\hour:%
  \ifnum \minute<10 0\fi%
  \number\minute%
  \ P.M.%
\fi%
}

\makeatother

\begin{document}

\baselineskip=18pt  
\numberwithin{equation}{section}  
\allowdisplaybreaks  


\thispagestyle{empty}

\vspace*{0.8cm} 
\begin{center}
  {\huge $\mathrm{Spin}(7)$-Manifolds as Generalized Connected Sums \\
 
  \smallskip
  
  and 3d $\mathcal{N}=1$ Theories}

 \vspace*{1.5cm}
{Andreas P. Braun and  Sakura Sch\"afer-Nameki}\\

 \vspace*{1cm} 
{\it  Mathematical Institute, University of Oxford \\
 Woodstock Road, Oxford, OX2 6GG, UK}\\
 {\tt maths.ox.ac.uk:andreas.braun, gmail:sakura.schafer.nameki}\\

\end{center}
\vspace*{.8cm}

\noindent
M-theory on compact eight-manifolds with $\mathrm{Spin}(7)$-holonomy is a framework for geometric engineering of 3d $\mathcal{N}=1$ gauge theories coupled to gravity. We propose a new construction of such $\mathrm{Spin}(7)$-manifolds, based on a generalized connected sum, where the building blocks are a Calabi-Yau four-fold and a $G_2$-holonomy manifold times a circle, respectively, which both asymptote to a Calabi-Yau three-fold times a cylinder. The generalized connected sum construction is first exemplified for Joyce orbifolds, and is then used to 
construct examples of new compact manifolds with $\mathrm{Spin}(7)$-holonomy. In instances when there is a K3-fibration of the $\mathrm{Spin}(7)$-manifold, we test the spectra using duality to heterotic on a $T^3$-fibered 
$G_2$-holonomy manifold, which are shown to be precisely the recently discovered twisted-connected sum constructions.

\newpage


\tableofcontents

\section{Introduction}
\label{sec:intro}

Geometric engineering of supersymmetric gauge theories is fairly well-understood in theories with at least four real supercharges. Most prominently in recent years, F-theory has established a precise dictionary between geometric data (and fluxes) and 6d and 4d theories with $\mathcal{N}=1$ supersymmetry, and dually, M-theory of course has a well-established dictionary between 3d and 5d theories and Calabi-Yau geometries. For 4d $\mathcal{N}=1$ the realization in terms of M-theory compatifications on $G_2$-holonomy manifolds is already much less well understood, in particular due to the scarcity of compact geometries of this type. Even less is known about M-theory on $\Spin(7)$-holonomy compactifications, which yield 3d $\mathcal{N}=1$ theories. 

The goal of this paper is to provide a new construction of compact $\Spin(7)$-holonomy manifolds, which have an interesting field theoretic counterpart in 3d. This is motivated on the one hand to build large sets of examples of 3d $\mathcal{N}=1$ vacua in M-theory/string theory, and we relate these constructions in certain instances to heterotic string theory on $G_2$-holonomy manifolds. A second motivation is to study $\Spin(7)$ compactification in the context of M/F-duality and uplifting these to supersymmetry breaking vacua in F-theory \cite{Vafa:1996xn} based on the observations that 3d $\mathcal{N}=1$ supersymmetry can be related to the circle-reduction of a not necessarily supersymmetric theory in 4d \cite{Witten:1994cga, Witten:1995rz}.

\begin{figure}
\begin{center}
\includegraphics[width=8.5cm]{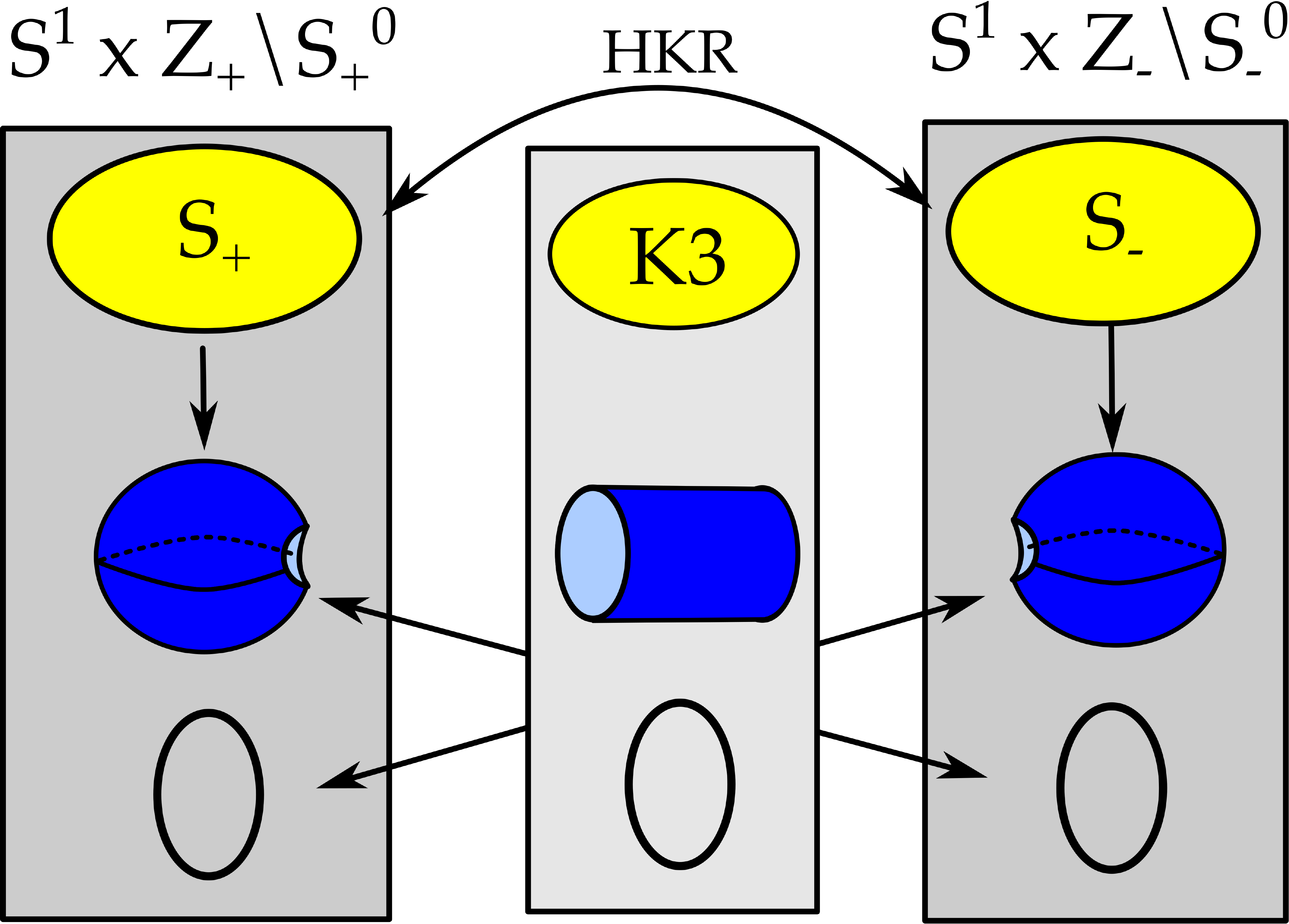}
\caption{The twisted connected sum (TCS) construction for $G_2$-manifolds: 
The two acyl Calabi-Yau three-fold building blocks are $X_\pm = Z_\pm \backslash S_\pm^0$, where $S_\pm$ denotes the K3-fibers of each building block. 
Unlike the $\Spin(7)$ case, here the gluing is done with a twist (or a Donaldson matching): a hyper-K\"ahler rotation (HKR) on the K3-fibers, and an exchange of the circles as shown above. The asymptotic region where the gluing is performed is K3$\times S^1 \times S^1 \times I$. 
Topologically the resulting $G_2$-manifold is K3-fibered over an $S^3$.\label{fig:TCSG2}}
\end{center}
\end{figure}

The construction that we propose, and refer to as \emph{generalized connected sum} (GCS) construction of $\Spin(7)$-manifolds is motivated by a recent development for compact $G_2$-holonomy manifolds in Mathematics, where a large class of compact $G_2$-holonomy manifolds were constructed using a \emph{twisted connected sum} (TCS) construction \cite{MR2024648, MR3109862, Corti:2012kd}. This construction relies on the decomposition of the $G_2$-holonomy manifold in terms of two asymptotically cylindrical (acyl) Calabi-Yau three-folds -- for a sketch see figure \ref{fig:TCSG2}. Following these mathematical developments, there  has been a resurgence in interest  in the string and M-theory compactifications, which for  TCS-manifolds have been studied in \cite{Halverson:2014tya, Halverson:2015vta, Braun:2016igl, Braun:2017ryx, Guio:2017zfn, Braun:2017uku,  Braun:2017csz, Braun:2018fdp}. For a review of M-theory on  $G_2$ and $\Spin(7)$-holonomy manifolds related to earlier results in the '90s and early '00s see \cite{Acharya:2004qe}.

One of the nice features of TCS $G_2$-manifolds is that the connected sum gives rise to a field theoretic decomposition in terms higher supersymmetric subsectors \cite{Guio:2017zfn, Braun:2017csz}: the asymptotic regions that are acyl Calabi-Yau three-folds (times a circle), give rise to 4d $\mathcal{N}=2$ subsectors, whereas the asymptotic neck region where the two building blocks are glued together is by itself K3 times a cylinder times $S^1$, and corresponds to a 4d  $\mathcal{N}=4$ subsector, which is present in the decoupling limit (infinite neck limit). The theory breaks to $\mathcal{N}=1$, when the asymptotic region is of finite size and the states of the $\mathcal{N}=4$ vector multiplet become massive, leaving only an $\mathcal{N}=1$ massless vector. This observation may be used to study non-perturbative corrections, e.g. M2-brane instantons \cite{Harvey:1999as}, which has been initiated for TCS-manifolds in \cite{Halverson:2014tya, Halverson:2015vta, Braun:2018fdp}.

For $\Spin(7)$-manifolds the type of constructions have thus far has been rather limited. There are (to our knowledge) two constructions of compact manifolds with $\Spin(7)$-holonomy due to Joyce: either in terms of a Joyce orbifold  $T^8/\Gamma$ \cite{joyce1996spin7} or a quotient by an anti-holomorphic involution of a Calabi-Yau four-fold ($\CY_4$) \cite{joyce1996spin7_new}. There are variations of these constructions using $\CY_4$ with $\Spin(7)$-necks in \cite{KovalevSpin7}. In both constructions, the quotient has singularities, which get resolved and shown to give rise to an eight-manifold with $\Spin(7)$-holonomy. Although it is relatively straight forward to construct string and M-theory compactifications on the Joyce orbifolds, understanding the second class of constructions is far less straight-forward. Recent attempts were made, in the context of M/F-theory duality for $\Spin(7)$-manifolds \cite{Vafa:1996xn}  in \cite{Bonetti:2013fma,Bonetti:2013nka}. One of the challenges in this is determining the 3d $\mathcal{N}=1$ theory, which is crucially related to the singular loci of the Calabi-Yau quotient, and not only the sector that is 3d $\mathcal{N}=2$. 

\begin{figure}
\begin{center}
\includegraphics[width=8cm]{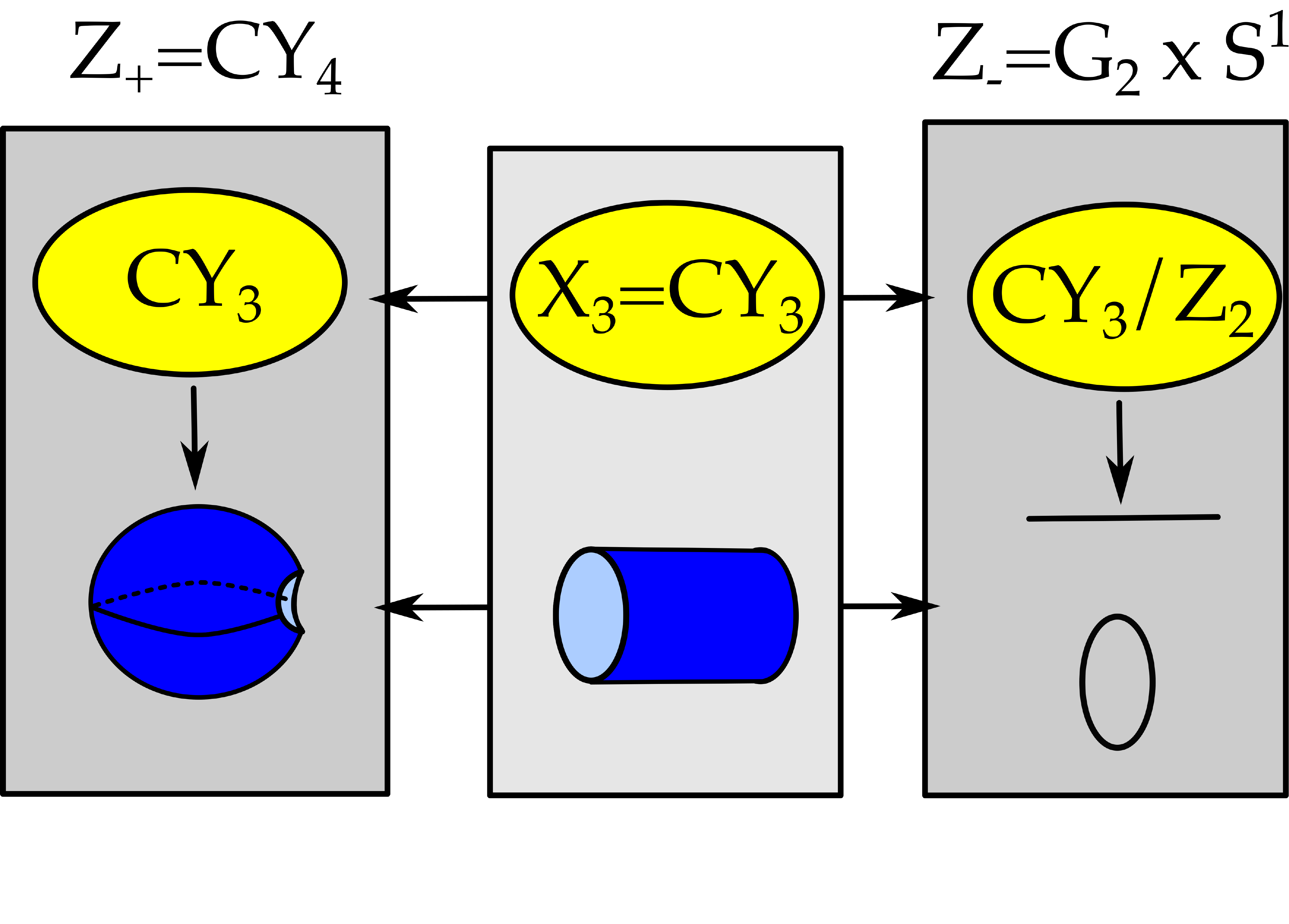}
\caption{The generalized connected sum (GCS) construction for $\Spin(7)$-manifolds proposed in this paper:
the left hand building block is an the acyl Calabi-Yau four-fold building block. We require this to be asymptotically approach a Calabi-Yau three-fold times a cylinder. One way to realize this is in terms of a $\CY_3$ fibration over an open $\mathbb{P}^1$. The right hand building block is a circle times a $G_2$-manifold, which asymptotes to a $\CY_3 \times I$ times a circle. The simplest way to realize this is in terms of $G_2 =(\CY_3 \times \R)/\Z_2$ where $\Z_2$ acts as an anti-holomorphic involution. 
The asymptotic region is $\CY_3 \times$cylinder. \label{fig:GCSSpin7}}
\end{center}
\end{figure}

In the present paper we provide a different approach to constructing $\Spin(7)$-manifolds, which is closer in spirit to the TCS construction. We will motivate a generalized connected sum (GCS)-construction where the building blocks are an open asymptotically cylindrical (acyl) Calabi-Yau four-fold ($\CY_4$) and a open asymptotically cylindrical (acyl) $G_2$-holonomy manifold (times $S^1$), which asymptote to a Calabi-Yau three-fold (times a cylinder/line). This is shown in figure \ref{fig:GCSSpin7}. Explicitly, we construct the acyl $\CY_4$ as
\be\label{CY34}
\CY_3 \hookrightarrow \CY_4 \rightarrow \mathbb{C} \,,
\ee
and the $G_2 = (\CY_3 \times \R)/\mathbb{Z}_2$, where $\Z_2$ acts as a reflection on the circle and anti-holomorphic involution on the $\CY_3$.  
In case the anti-holomorphic involution acts freely, these are so-called `barely $G_2$-manifolds' \cite{Harvey:1999as}, which have holonomy group $SU(3) \rtimes \mathbb{Z}_2$.

Field theoretically we will show that each of the building blocks will give rise to a 3d $\mathcal{N}=2$ subsector, and the asymptotic region $\CY_3 \times$ cylinder results in the limit of an infinite neck region in a 3d $\mathcal{N}=4$ sector. At finite distance, some of the modes of the vector multiplet of 3d $\mathcal{N}=4$ become massive and the theory is broken to 3d $\mathcal{N}=1$.

\begin{figure}
\begin{center}
\includegraphics[width=6cm]{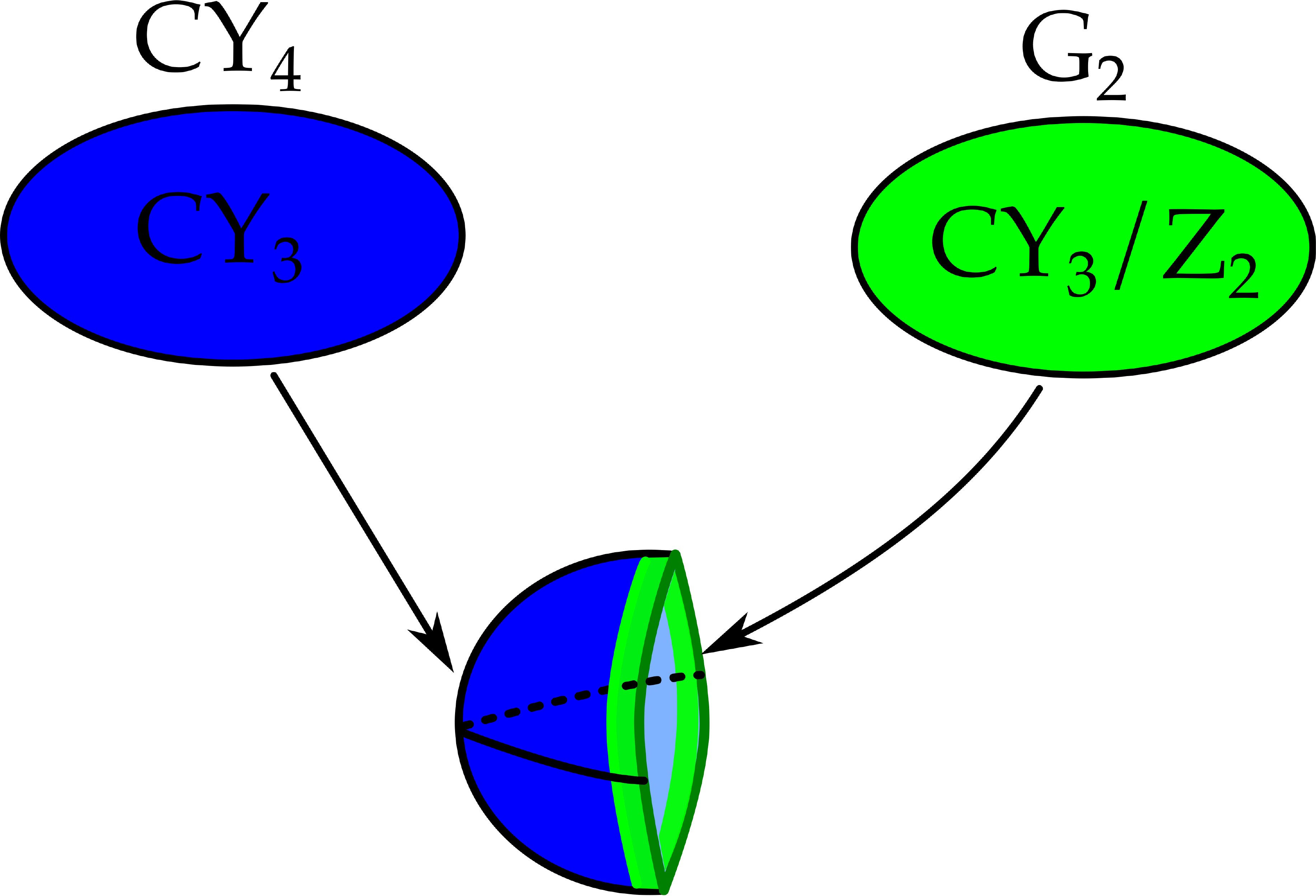}
\caption{GCS-construction of $\Spin(7)$-manifolds, from the point of view of a Calabi-Yau four-fold quotient by an anti-holomorphic involution. The $\CY_4$ is as in (\ref{CY34}) and the picture shows the geometry after the quotient. Away from the boundary circle, which is the fixed locus of the involution, we have a $\CY_3$ fibered over a disc, which is locally a $\CY_4$. Above the boundary circle (dark green) the quotient acts non-trivially on the $\CY_3 \times$ interval, shown in light green, which results in a $G_2$. Pulling the base apart, results in the  GCS-decomposition of the resulting $\Spin(7)$-manifold.
\label{fig:CY4Pic}}
\end{center}
\end{figure}

We motivate this construction in section \ref{sec:JoyceExample} by considering a simple $\Spin(7)$ Joyce orbifold, $T^8/\Gamma$, which is known to have a related $G_2$ Joyce orbifold, where the precise connection is through M-theory on K3 to Heterotic on $T^3$ duality applied to this toroidal setting \cite{Acharya:1996ef}. It is known that such Joyce orbifolds have a TCS-decomposition \cite{Braun:2017uku}, which we apply to the $G_2$ Joyce orbifold, and subsequently uplift the decomposition to the $\Spin(7)$ Joyce orbifold. The TCS-building blocks then map precisely to an open  $\CY_4$ and $G_2\times S^1$, respectively. This motivates our  general construction of GCS $\Spin(7)$-manifolds, which will be given in section \ref{sect:generalconstruction}. 

One may ask whether these constructions in fact globally fit together to (resolutions of) $\CY_4$ quotients, like in Joyce's constructions in \cite{joyce1996spin7_new}. Whenever the acyl $G_2$ manifold is realized as (the resolution of) a quotient of $X_3 \times \R$, this is indeed the case and we can write our GCS $\Spin(7)$-manifolds globally as resolutions of quotients of $\CY_4$ by an anti-holomorphic involution. Here, one starts with a Calabi-Yau four-fold fibered by a Calabi-Yau three-fold $\CY_3$ as in (\ref{CY34}), and acts on it with an anti-holomorphic involution on the $\CY_3$-fiber combined with an action on the base $\mathbb{P}^1$ given by $z_i \leftrightarrow -\bar{z}_j$, $i\not= j$, where $z_i$ are the homogeneous coordinates. The fixed locus of this on the base is a circle and the quotient space is half-$\CY_4$, which is a $\CY_3$-fold fibered over a disc, and {in the vicinity of the boundary circle we obtain $(\CY_3 \times \mathbb{R})/\mathbb{Z}_2$ which is a (barely) $G_2$-manifold, times an $\bbS^1$}. Pulling the base half sphere apart, gives the decomposition into the GCS building blocks. This is sketched in figure \ref{fig:CY4Pic}. The key difference to the examples and constructions in \cite{joyce1996spin7_new} is however that the singularities are either absent or occur over a locus of real dimension at least 1. In Joyce's construction via $\CY_4/\sigma$ the singularities are only point-like and get resolved using an ALE-space. It seems likely that the analog to this procedure in our construction is gluing in the open, acyl $G_2$-manifold.

The paper is structured as follows: we begin with some background material on $\Spin(7)$-holonomy and M-theory compactifications to 3d $\mathcal{N}=1$ in section \ref{sec:Spin7M}. In section \ref{sec:JoyceExample} we motivate our construction by first considering a Joyce orbifold example. Section \ref{sect:generalconstruction} contains our proposal for the general construction of GCS $\Spin(7)$-manifolds. We also comment on how this construction relates to and differs from other known setups and provide an example of a new $\Spin(7)$-manifold. 
For some examples of GCS constructions there is a K3-fibration of the $\Spin(7)$-manifold, and we utilize this to apply M-theory/heterotic duality and construct the dual heterotic on $G_2$ compactifications in section \ref{sec:MHet}, and match the spectra of the theories. 
We conclude with a discussion and outlook in section \ref{sec:Outlook}. For reference we include a brief summary of the TCS-construction of $G_2$-manifolds in appendix \ref{app:TCS}.


\section{$\Spin(7)$-Holonomy, M-theory and 3d $\mathcal{N}=1$ Theories}
\label{sec:Spin7M}

As a preparation for this work and to introduce some notation, this section reviews some basic aspects of manifolds with holonomy group $\Spin(7)$, see \cite{joyce2000compact} for a detailed discussion.

\subsection{$\Spin(7)$-Holonomy}

A compact orientable 8-dimensional manifold $Z$ which has a Ricci-flat metric $g$ with holonomy group contained in $\Spin(7)$ supports a closed, self-dual four-form $\Psi = \ast_g \Psi$, which can be expressed in local coordinates (in which $g$ is the euclidean metric) as
\begin{equation}
\begin{aligned}
\Psi =\,\,&  dx_{1234} + dx_{1256} + dx_{1278} +dx_{1357} -dx_{1368}-dx_{1458}-dx_{1467} \\
&-dx_{2358}-dx_{2367}-dx_{2457}+dx_{2468}+dx_{3456}+dx_{3478}+dx_{5678} \, .
\end{aligned}
\end{equation}
Here, we use $dx_{ijkl}$ as a shorthand for $dx_i\wedge dx_j\wedge dx_k \wedge dx_l$. This four-form is in the stabilizer of the action of the holonomy group $\Spin(7)$. Such a form defines a $\Spin(7)$ structure, which is called torsion free if $\Psi$ is closed and self-dual. 

Having this structure in place does not necessarily mean that the holonomy group is exactly $\Spin(7)$ and, as long as $Z$ is simply connected, we may discriminate between different cases by computing the $\hat{A}$ genus, which tells us the number of covariantly constant spinors:
\begin{equation}
\begin{array}{c|c}
\hat{A} &  \hbox{hol}(g) \\
\hline
 1 & \Spin(7) \\
 2 & SU(4) \\
 3 & Sp(2) \\
 4 & SU(2) \times SU(2) \\
\end{array}
\end{equation}
All of these manifolds have $b^1(Z)=0$ and the remaining independent Betti numbers are related to the $\hat{A}$ genus by 
\begin{equation}\label{Aroof}
24 \hat{A} = -1 -b^2 + b^3 +b^4_+ - 2b^4_- \,,
\end{equation}
where $b^4_\pm$ are the dimensions of the (anti)self-dual subspaces of $H^4(Z,\Q)$. As we are interested in compactifications of M-theory which preserve $\mathcal{N}=1$ supersymmetry in three dimensions, we will only be interested in the case $\hat{A}=1$. 

For eight-manifolds $Z$ with holonomy contained in $\Spin(7)$ and $\hat{A}(Z) =1$, a necessary and sufficient condition for the holonomy group to be all of $\Spin(7)$ is that $Z$ is simply connected, $\pi_1(Z)=0$ \cite{joyce1996spin7,joyce2000compact}. This still allows for cases with non-trivial subgroups of $\Spin(7)$, and we will see examples of such spaces later on. We will refer to manifolds $Z$ with $\hat{A}(Z) =1$ and a metric $g$ with $\hbox{hol}(g) \subseteq$ $\Spin(7)$ as barely $\Spin(7)$-manifolds.

The dimension of the moduli space of Ricci-flat metrics on a $\Spin(7)$-manifold is given by $b^4_-+1$. Together with $\hat{A}=1$, this number is already determined by the Euler characteristic and the two Betti numbers $b_2$ and $b_3$ by using \eqref{Aroof}
\begin{equation}\label{eq:b4-intermsofchib2b3}
b^4_- =  {\chi\over 3} - 9 - b_2 + b_3 \, .
\end{equation}

Calibrated submanifolds of $\Spin(7)$-manifolds must be of real dimension four and are called Cayley submanifolds. The dimension of the moduli space of such a Cayley submanifold $N$ is \cite{Mclean96,joyce2000compact}
\begin{equation}
\begin{aligned}
m_N =  &\tau(N) -\tfrac12 \chi(N) -\tfrac12 N \cdot N \, ,
\end{aligned}
\end{equation}
where $\tau(N)$ is the Hirzebruch signature and $\chi(N)$ is the Euler characteristic. Note that this expression evaluations to 
\begin{equation}
m_{N} =  4 -\tfrac12 N \cdot N
\end{equation}
for a Cayley submanifold which is a K3 surface, where $\tau(N) = 16$ and $\chi(N) = 24$. It hence seems sensible to assume there exist $\Spin(7)$-manifolds fibered by K3 surfaces over a four-dimensional base. Using duality to heterotic strings, it is precisely the existence of such fibrations which we will conjecture and exploit. In contrast, note that for a Cayley submanifold with the topology of a four-dimensional torus $T^4$, the same computation gives
\begin{equation}
\begin{aligned}
m_N  =&  - \tfrac12 N \cdot N 
\end{aligned}
\end{equation}
so that $\Spin(7)$-manifolds cannot possibly be fibered by calibrated four-tori.

\subsection{M-theory on $\Spin(7)$-Manifolds}

M-theory on a $\Spin(7)$-manifold gives rise to 3d $\mathcal{N}=1$ supersymmetric theory and the spectrum is encoded in the topological data of the $\Spin(7)$-manifold as follows:
\be
\begin{array}{c|c|c}
\# & \hbox{3d $\mathcal{N}=1$ Multiplet} & \hbox{11d Origin}\cr \hline
b_3(Z) &\hbox{Scalar }   & C_3 \supset \sum_{i=1}^{b_3} \rho^{(3)}_i \varphi^i\cr 
b_2 (Z) &\hbox{Vector }  & C_3 \supset \sum_{i=1}^{b_2} \omega^{(2)}_\alpha v_\mu^\alpha  \cr 
b_4^-(Z) & \hbox{Scalar} &\delta g_{\mu\nu} \supset  \sum_{I=1}^{b_4^-}(\xi_I^{(4)})_{\mu\rho\sigma\tau} \Phi_\nu\,^{\rho\sigma\tau} \delta \varphi^I \cr 
1 & \hbox{Scalar} & \hbox{Volume modulus} 
\end{array}\ee
where $\omega^{(2)}$, $\rho^{(3)}$ are a basis of harmonic two and three-forms, and $\xi_I^{(4)}$ are a basis of harmonic anti-self-dual four-forms. For 3d $\mathcal{N}=1$ the scalar multiplet has only a real scalar as its bosonic component and the vector multiplet is just a 3d vector. As we can dualize the 3d vectors to real scalars, compactifications of M-theory on $\Spin(7)$-manifolds hence give rise to
\begin{equation}\label{eq:specMspin7}
n_s = b^4_- + 1 + b^2 + b^3 =  {\chi(Z) \over 3} - 8 + 2 b^3  
\end{equation}
massless real scalars at the classical level. Anomaly cancellation furthermore requires the introduction of 
\begin{equation}
 N_{M2} = \frac{1}{24} \chi(Z)- {1\over 2}\int_Z G_4 \wedge G_4  
\end{equation}
space-time filling M2-branes \cite{Vafa:1995fj,Sethi:1996es}. As such M2-branes can freely move on $Z$, they each contribute a further 8 real degrees of freedom. 

In the absence of $G_4$-flux, the effective action for a smooth $\Spin(7)$-manifold is a 3d $\mathcal{N}=1$ field theory with $b_2$ abelian vectors, and scalars with the following kinetic terms 
\be
\mathcal{G}_{I J}  \sim {\int_Z \xi^{(4)}_I \wedge \xi^{(4)}_J} \,,\qquad 
\mathcal{M}_{\alpha\beta} \sim \int_Z \omega^{(2)}_\alpha \wedge \star \omega^{(2)}_\beta \, .
\ee
The theory for a smooth $\Spin(7)$-manifold is an abelian gauge theory. With singularities this can enhance to non-abelian gauge symmetries.
With $G_4$-flux, additional Chern-Simons terms and scalar potential are generated -- for an in depth discussion of the effective theory see \cite{Papadopoulos:1995da, Becker:2000jc, Gukov:2001hf, Acharya:2002vs, Becker:2003wb, Tsimpis:2005kj, Prins:2013wza, Bonetti:2013fma}. In this paper we will not consider fluxes but focus on the geometric constructions. Of course it is interesting to consider this in the future and study the effects of these on supersymmetry breaking and potential obstructions to dualities, both M/F-duality \cite{Prins:2015nda}, where in particular 4d Poincar\'e invariance could be broken, as well as obstructions to  M-theory/weakly-coupled heterotic duality \cite{Melnikov:2017wcf, Melnikov:2017yvz}.

\section{A $\Spin(7)$ Joyce Orbifold as a Generalized Connected Sum}
\label{sec:JoyceExample}

\subsection{Setup and Motivation}
\label{sec:Setup}

Our  goal is to construct new classes of $\Spin(7)$-manifolds -- which we have motivated from various points of view in the introduction.  
The construction which we will end up with is inspired by combining two ideas: 
\begin{enumerate}
\item The recent construction of $G_2$-holonomy manifolds as twisted connected sums (TCS), with each building block an asymptotically cylindrical Calabi-Yau three-fold. We review this construction in appendix \ref{app:TCS}.
\item M-theory on K3/heterotic on $T^3$ duality.
\end{enumerate}

Combining these ideas will lead us to consider a generalized connected sum (GCS)  construction, where -- as we will show -- the building blocks are acyl CY four-folds and $G_2$-manifolds, respectively. To motive this we start with a well-known construction by Joyce of both $G_2$ and $\Spin(7)$-manifolds and a well-known duality, between M-theory and heterotic strings, which will be discussed in more detail in section
 \ref{sec:MHetDual}, between 
\be
\hbox{M-theory on K3}  \quad \longleftrightarrow \quad \hbox{Heterotic on $T^3$}
\ee
This duality in 7d is based on the agreement of the moduli spaces of the two compactifications
\be
\mathcal{M}_{het/T^3} = \mathcal{M}_{M/K3} = \left[ SO(3, \mathbb{Z}) \times SO(10, \mathbb{Z})\right] \backslash SO(3,19)/\left[SO(3) \times SO(10)\right] \,  \times \mathbb{R}^+ \,,
\ee
which is both the Narain moduli space of heterotic string theory on $T^3$ and the moduli space of Einstein metrics on K3 \cite{Witten:1995ex} (for a recent exposition in the context of fiber-wise application see \cite{Braun:2017uku}). The string coupling on the heterotic side is matched with the volume modulus of the K3.
We can fiber this over a four-manifold $M_4$ in such a way that there is a duality of 3d $\N=1$ theories
\be
\hbox{M-theory on $\Spin(7)$-manifold $Z_8$}  \quad \longleftrightarrow \quad \hbox{Heterotic on $G_2$-manifold $J_7$}
\ee
where $Z_8$ is K3-fibered over $M_4$ and $J_7$ has $T^3$-fibers. This duality has been tested in the case when both manifolds are Joyce orbifolds in \cite{Acharya:1996ef}. 
More generally, fiberwise application of this duality will lead us to a correspondence between two realizations of a 3d $\mathcal{N}=1$ theory, in terms of a $\Spin(7)$-compactification of M-theory, and a $T^3$-fibered $G_2$-holonomy manifold\footnote{The existence of $T^3$-fibrations of $G_2$ manifolds has also been conjectured in the context of mirror symmetry for $G_2$ manifolds in \cite{Braun:2017csz}.}. It is this setup which will motivate our construction of GCS $\Spin(7)$-manifolds: it is the analog of the TCS-decomposition for $G_2$-manifolds on the heterotic side, mapped to M-theory using the duality. In this way we obtain a dual pair of connected sums:
\be
\hbox{M-theory on GCS $\Spin(7)$-manifold} \quad \longleftrightarrow \quad \hbox{Heterotic on TCS $G_2$-manifold}
\ee
As a warmup we now show how this works for a simple $G_2$ Joyce orbifold, which has a TCS-decomposition 
and determine what this decomposition corresponds to on the M-theory side. This will give a first hint as to what the general connected sum construction will be for $\Spin(7)$-manifolds.

\subsection{Joyce Orbifolds: $\Spin(7)$- and $G_2$-Holonomy}

We will start the construction of a $\Spin(7)$-manifold $Z_8$ by considering a Joyce orbifold  $T^8/\Gamma$, where $\Gamma = \mathbb{Z}_2^4$, where each generator of the order two subgroups acts as follows \cite{joyce2000compact}: 
\be\label{eq:orbispin7}
\begin{array}{|c||c|c|c||c||c|c|c|c|c|c|c|}\hline
& x_1 & x_2 & x_3 & x_4 & x_5 & x_6 & x_7 & x_8 \cr  \hline
\alpha &  - & -&-&-& +&+&+&+  \cr \hline\hline
\beta&  +&+ &+&+&-&-&-&-\cr \hline
\gamma &  {1\over 2}- &  {1\over 2}-  & + &+&    {1\over 2}-&  {1\over 2}-  &+ & +\cr \hline
\delta &     -  & + &   {1\over 2} -& + &  {1\over 2} - & +  & {1\over 2} -& + \cr  \hline
\end{array}
\ee
The entries ${1\over 2}-$ denote $x \rightarrow - x + {1\over 2}$. 
The singular sets are locally given by 
\be
\ba
S_\alpha:  &\quad 4 \times \C^2/\Z_2 \times T^4/\{\pm 1\} \cong  \C^2/\Z_2 \times  K3 \cr 
S_\beta:  &\quad 4 \times  \C^2/\Z_2  \times T^4/\{\pm 1\} \cong  \C^2/\Z_2 \times K3 \cr 
S_\gamma:  &\quad 2 \times  \C^2/\Z_2 \times T^4  \cr 
S_\delta:  &\quad 2 \times  \C^2/\Z_2 \times T^4  \,,
\ea\ee
as well as $64$ fixed points of $\alpha\beta$, the singularities of which are modelled on $\C^4/\Z_2$. It was shown by Joyce that the resolution of the orbifold in this way yields a $\Spin(7)$-holonomy manifold.

From this, the Betti numbers are computed as follows. First note that the only even classes in $H^\bullet(T^8)$ under the $\Z_2^4$ (besides $H^0(T^8)$ and $H^8(T^8)$) are given by $14$ classes in $H^4(T^8)$. Resolving the $8$ singularities of the form $\C^2/\Z_2 \times T^4/\{\pm 1\}$ gives
\begin{equation}
\begin{array}{c|c|c}
 &  \C^2/\Z_2  \times T^4/\{\pm 1\}  &  \widetilde{\C^2/\Z_2}  \times T^4/\{\pm 1\}  \\\hline
b^2 & 6 & 7 \\
b^3 & 0 & 0 \\
b^4 & 1 & 7 
\end{array}
\end{equation}
whereas the $4$ singularities of the form $\C^2/\Z_2 \times T^4$ yield
\begin{equation}
\begin{array}{c|c|c}
 &  \C^2/\Z_2  \times T^4 &  \widetilde{\C^2/\Z_2}  \times T^4  \\\hline
b^2 & 6 & 7 \\
b^3 & 4 & 8 \\
b^4 & 1 & 7 
\end{array}
\end{equation}
Hence each of these $12$ singularities contributes $b^2 = 1$ and $b^4 = 6$ and only the second $4$ contribute a non-zero $b^3 = 4$. Finally, the resolution of the $64$ singularities of the form $\C^4/\Z_2^2$ contribute $b^4=1$ and nothing else. Altogether we can compute 
\begin{equation}
\begin{aligned}
b^2 &= 0 + 8 \cdot 1 + 4\cdot 1 +0  = 12 \\
b^3 &= 0 + 4 \cdot 4 +0 = 16 \\
b^4 &= 14 + 8 \cdot 6 + 4\cdot 6 + 64 = 150 \,.
\end{aligned}
\end{equation}

The key observation to make is that 
this $\Spin(7)$-manifold gives rise naturally to a $G_2$-manifold as follows: consider the $T^7$ given by $x_1, x_2, x_3, x_5, x_6, x_7, x_8$, and act with $\mathbb{Z}_2^\beta \oplus \mathbb{Z}_2^\gamma \oplus \mathbb{Z}_2^\delta$. This is a Joyce construction of $G_2$-manifolds by orbifolds as already observed in \cite{Acharya:1996ef}.

\subsection{TCS-decomposition of the Joyce $G_2$-Manifold}

\begin{figure}
\begin{center}
  \includegraphics[height=9cm]{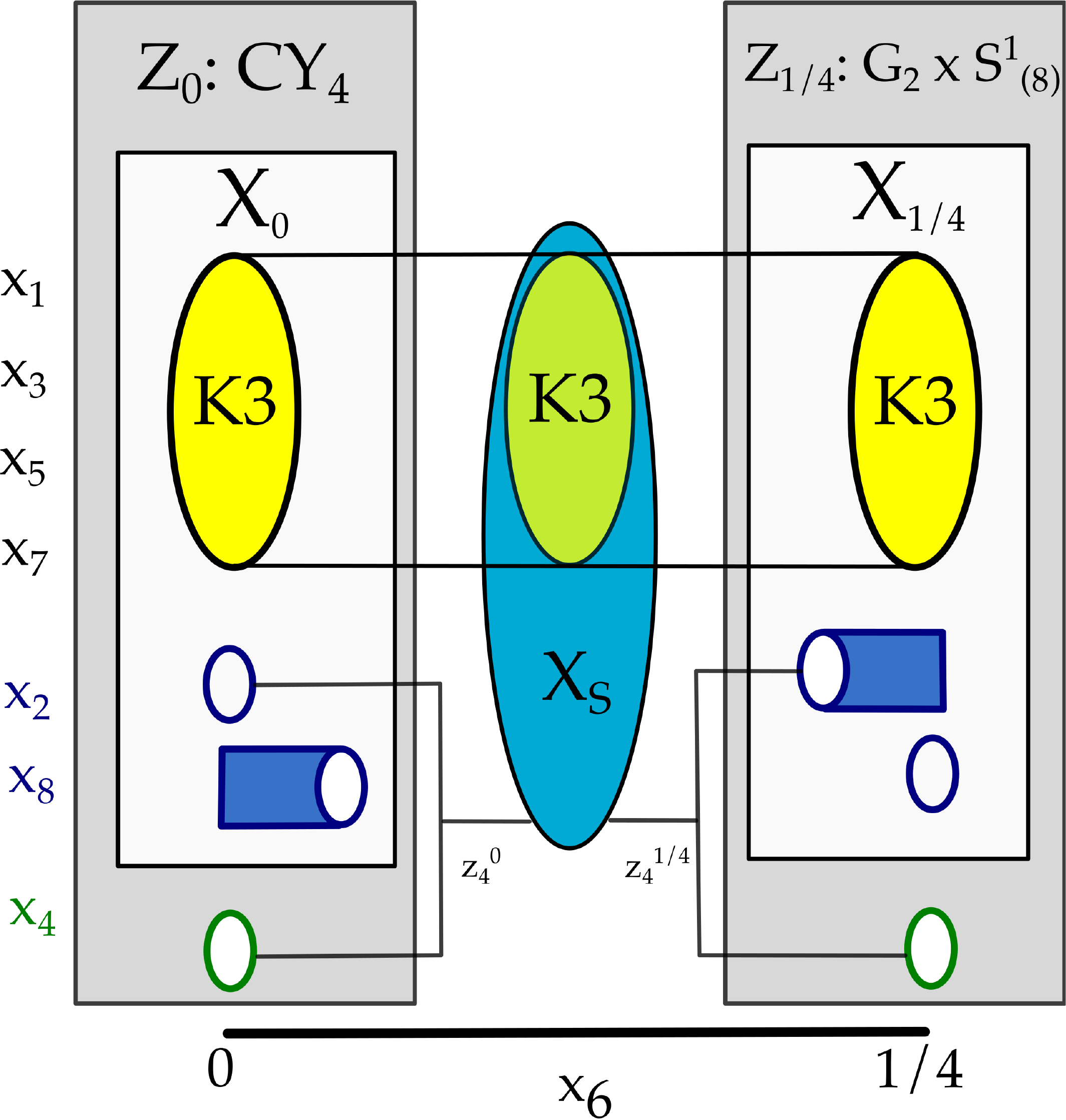}
 \caption{\label{fig:JoyceG2TCS} The GCS decomposition of the $\Spin(7)$ Joyce orbifold: The building blocks are $Z_0$ and $Z_{1/4}$, which are an open Calabi-Yau four-fold and an open $G_2$-manifold times $S^1_{(8)}$. These are glued together along the $x_6$ coordinate. Both asymptote to $X_s$ times a cylinder, where $X_s$ is the Schoen Calabi-Yau three-fold. In this particular example each building block contains a K3, which is also present in the neck region of the Schoen Calabi-Yau.}
\end{center}
\end{figure}

Our goal is to first identify the TCS description of this Joyce $G_2$-orbifold and to then lift this to a connected sum description, GCS, for the $\Spin(7)$-manifold. 
First note that $T^7/\langle \beta, \gamma, \delta\rangle$ is fibered over an interval 
\be\label{x6Fib}
x_6: (\beta, \gamma, \delta) = (-, -{1\over 2}, +) 
\ee
which identifies this as $x_6 \sim -x_6 \sim -x_6 + {1\over 2}$, and thus $x_6 \in I = [0, 1/4]$. 
The generator $\delta$ is the stabilizer of $x_6$ and so 
\be
T^7 /\langle \delta\rangle = T^3_{(2, 6, 8)} \times T_{(1,3,5,7)}^4/{\mathbb{Z}_2^\delta}   =  T^3 \times \mathrm{K3}\,.
\ee
The notation $T^3_{(2,6,8)}$ indicates the three-torus along the coordinates $x_2, x_6, x_8$. 
We now need to identify the action of $\beta$ and $\gamma$ on this space. 
Again, there will be two components, which we denote by $X_0$ and $X_{1/4}$, located at $x_6=0, 1/4$, which are given by
\be
\ba
X_0:\qquad   & S^1_{(8)}\times \left(\left.T^4_{(1,3,5,7)}/ \mathbb{Z}_2^\delta \times S^1_{(2)} \times I_{x^6} \right)\right/ \mathbb{Z}_2^\beta \cr 
X_{1/4}:\qquad & S^1_{(2)}\times \left(\left.T^4_{(1,3,5,7)}/ \mathbb{Z}_2^\delta \times S^1_{(8)} \times I_{x^6} \right)\right/ \mathbb{Z}_2^\gamma\,.
\ea
\ee
Each of these halves is an open, K3-fibered CY three-fold. To see this introduce the coordinates 
\be
X_0:\quad \left\{
\ba
z_1^0 & = x_1 + i x_3 \cr 
z_2^0 & = x_5 + i x_7 \cr 
z_3^0 & = x_6 + i x_8 
\ea\right. 
\qquad 
X_{1/4}:\quad \left\{
\ba
z_1^{1/4} & = x_7 + i x_3 \cr 
z_2^{1/4} & = x_5 + i x_1 \cr 
z_3^{1/4} & = x_6 + i x_2  
\ea\right. 
\ee
The K3s are along $z_1, z_2$, where $\delta$ acts non-trivially. 
The remaining coordinates combine with the interval coordinate $x_6$. At $x_6=0 (1/4)$ the $x_8 (x_2)$ circle pinches. A sketch is shown in figure \ref{fig:JoyceG2TCS}. The holomorphic three-form is invariant under these actions, making these open $CY_3$.

\subsection{Uplift to GCS-decomposition of the $\Spin (7)$-Manifold}\label{sect:firstspin7asTCS}

Adding back the circle along the $x_4$ coordinate, as well as the additional orbifold generator $\alpha$, $\alpha$ acts trivially on $x_6$, so that the fibration over the interval remains intact, but $\alpha: z_1^0 \rightarrow  -z_1^0$, so that the $\Omega^{(3,0)}$ form is not invariant any longer. Let us now again define two building blocks along the $x_6$ interval, close to either boundary. The interesting observation is that the two halves behave quite differently: first consider $x_6 =0$. Define another complex coordinate with the action of $\alpha$
\be
z_4^0 \equiv x_2 + i x_4  \  \stackrel{\alpha}{\longrightarrow}  \ - z_4^0 \,.
\ee
Together with the action of $\alpha$ on $z_i^0$, we see that $\Omega^{(4,0)}_0 = d z_1^0 \wedge d z_2^0 \wedge d z_3^0 \wedge d z_4^0$ is invariant, and we obtain a building block which is an open Calabi-Yau four-fold 
\be
Z_0 :\qquad  \left.\left(X_0 \times S^1_{(2)} \times S^1_{(4)} \right)\right/ \mathbb{Z}_2^{\alpha} \,.
\ee
At the other end of the $x_6$ interval we have 
\be
Z_{1/4}:\qquad \left.\left( X_{1/4} \times S^1_{(8)} \times S^1_{(4)}\right) \right/\mathbb{Z}_{2}^\alpha 
= S^1_{(8)} \times M_7\,,
\ee
where $M_7$ will be shown to be a $G_2$ holonomy manifold. The action of $\alpha$ is 
\be
\alpha:
\qquad\left\{ 
\ba
\Omega^{(3,0)}_{1/4} & \  \rightarrow \ \overline{\Omega^{(3,0)}_{1/4}} \cr 
x_4 & \ \rightarrow \ - x_4 \,.
\ea\right.
\ee
From this we can define the $G_2$-form
\be
\Phi_{3} = dx_4 \wedge \omega^{(1,1)} + \Re \Omega^{(3,0)}\,,\qquad 
\ast \Phi_3 = {1\over 2} \omega^{(1,1)}\wedge \omega^{(1,1)} - dx_4 \wedge \Im \Omega^{(3,0)}\,,
\ee
which is invariant, as follows from the action of $\alpha$ and the coordinates $z^{1/4}$. 

In summary we obtained a GCS-construction of a $\Spin(7)$ Joyce orbifold $Z$, which has two building blocks, an open $\CY_4$ and an open $G_2$, $W$ times a circle, respectively. The geometry in the middle of the $x_6$ internal is 
\be
Q = X_s \times S^1_{(8)} \times I_{(6)}\,,
\ee 
where the $X_s$ is a $\CY_3$ along the directions $1,2,3,4,5,7$. 
Not too surprisingly, this is in fact the Schoen Calabi-Yau three-fold. The action along these coordinates, written in terms of the complex coordinates of $Z_0$ is
\be\ba
\alpha:\qquad  (z_1^0, z_2^0, z_4^0)& \quad \rightarrow \quad (-z_1^0, z_2^0, - z_4^0)   \cr 
\delta:\qquad (z_1^0, z_2^0, z_4^0 )& \quad \rightarrow \quad \left(-z_1^0 +i {1\over 2}, -z_2^0, z_4^0 \right) \,.
\ea\ee
This is precisely the action required to get the Schoen Calabi-Yau as a Joyce orbifold (see \cite{Braun:2017uku} for a detailed discussion of that). Inside $X_s$, there is a K3 surface along $z_1^0, z_2^0$. On the other hand the $\CY_4$ $Z_0$ is fibered by $X_s$ over $x_6, x_8$.

Note that we may also think about this Spin(7) orbifold as a quotient of a Calabi-Yau four-fold $X_4$ by an anti-holomorphic involution. The action of $\alpha, \beta, \delta$ respects the holomorphic coordinates $z_i^0$ on $T^8$ and leaves $\Omega^{(4,0)}_0$ invariant, so that it produces a Calabi-Yau orbifold. The full Spin(7) orbifold is then formed by acting with $\gamma$, which acts as an anti-holomorphic involution. We expect this structure to persist after resolution.

The Joyce orbifold (\ref{eq:orbispin7}) has several generalizations, which can be parametrized in terms of the following data:
\be\label{JoyceSpin7}
\begin{array}{|c||c|c||c||c|c|c|c|c|c|c|c|}\hline
& x_1 & x_2 & x_3 & x_4 & x_5 & x_6 & x_7 & x_8 \cr  \hline
\alpha &  - & -&-&-& +&+&+&+  \cr \hline
\beta&  +&+ &+&+&-&-&-&-\cr \hline
\gamma &  c_1 - &  c_2-  & + &+&    c_5-&  c_6- &+ & +\cr \hline
\delta &     d_1 -  & + &   d_3 - & + & d_5-  & +  & d_7- & + \cr  \hline
\end{array}
\ee
where the shift vectors can take the following values
\be
\ba
I: & \qquad {\bf c}  = {1\over 2} (1,1,1,1) \,,\qquad {\bf d} = {1\over 2} (0, 1,1,1) \cr 
II: &\qquad  {\bf c}  = {1\over 2} (1,0,1,0) \,,\qquad {\bf d} = {1\over 2} (0, 1,1,1) \cr 
III: & \qquad {\bf c}  = {1\over 2} (1,1,1,0) \,,\qquad {\bf d} = {1\over 2} (0, 1,1,0) \cr 
IV: &\qquad  {\bf c}  = {1\over 2} (1,0,1,0) \,,\qquad {\bf d} = {1\over 2} (0, 1,1,0) \,.
\ea
\ee
Case $I$ is the orbifold studied above. Instead of the fibration over the interval (\ref{x6Fib}), in this general setting, we  consider pulling these orbifolds apart along $x_3$, with the interval $x_3 \in [0, 1/4]$. There is an asymptotic middle-region, where the space is 
$X_s = T^{6}_{(1,2,5,6,7,8)} / \langle \beta, \gamma \rangle$, which in fact is again the Schoen Calabi-Yau three-fold for all of these Joyce orbifolds. The asymptotic manifolds at the fixed points $x_3=0, {1\over 4}$ are again open a Calabi-Yau four-fold and $G_2$-manifold $M_7$, respectively, given by
\be\ba
Z_0: \qquad &   (I_{(3)}\times S^1_{(4)}\times X_0)/\mathbb{Z}_2^\alpha \cr
Z_{1/4}:\qquad & S^1_{(4)} \times  (I_{{(3)}} \times X_0)/\mathbb{Z}_2^\delta \,.
\ea\ee
The analysis will follow very much along the lines of the first example we studied and we will instead now move to generalize this construction beyond Joyce orbifolds.

We have seen that the Joyce orbifold $Z_8$ (\ref{eq:orbispin7}) has a natural decomposition in terms of a GCS-construction, where one building block is an open Calabi-Yau four-fold, and the other is a $G_2$-manifold times a circle, where both geometries asymptote to a Calabi-Yau three-fold $X_s$ times a cylinder. With this example in mind, we now turn to providing a generalization of this construction in the next section.


\section{$\Spin(7)$-Manifolds as Generalized Connected Sums}
\label{sect:generalconstruction}

\subsection{The Construction}

In the discussion of the last section, we were led to think about $\Spin(7)$ Joyce orbifolds as being glued from two eight-manifolds: a  non-compact Calabi-Yau four-fold and the product of a non-compact $G_2$-manifold with a circle. These two halves are glued along their overlap, which is the product of a Calabi-Yau three-fold with an open cylinder $I\times \bbS^1$ -- see figure \ref{fig:GCSSpin7}. We are now going to propose a generalization of this procedure in which both the two building blocks $Z_\pm$ and the resulting $\Spin(7)$-manifold $Z$ are no longer (resolutions of) orbifolds. To start this discussion, let us first propose the following definitions: 

\subsubsection*{Definition:}
\emph{An asymptotically cylindrical (acyl) Calabi-Yau four-fold $Z_+$ is a non-compact algebraic four-fold, which admits a Ricci-flat metric, is simply connected, and is diffeomorphic (as a real manifold) to the product of a cylinder $I\times\bbS^1$ and a compact Calabi-Yau three-fold $X_3$ outside of a compact submanifold $\kappa$. The Ricci-flat metric of $Z_+$ exponentially asymptotes to the Ricci flat metric on the product $I\times \bbS^1 \times X_3 \cong X_4\setminus \kappa$.}
\\

Asymptotically Calabi-Yau manifolds were discussed in \cite{Haskins2015}. Similar to the explicit construction of acyl three-folds in \cite{MR3109862,Corti:2012kd}, we expect to able to construct $Z_+$ by excising a fiber $X_3$ from a compact four-fold $\tilde{Z}_+$ with $c_1(\tilde{Z}_+) = [X_3]$, which is fibered by Calabi-Yau three-folds. This implies that we can think of producing $Z_+$ by appropriately cutting a compact, $\CY_3$-fibered Calabi-Yau four-fold in half, or via toric methods as in \cite{Braun:2016igl}. Likewise we define: 

\subsubsection*{Definition:}
\emph{An asymptotically Calabi-Yau (acyl) $G_2$-manifold $Z_-$ is a non-compact manifold of $G_2$ holonomy, which is diffeomorphic to the product of an interval $I$ with a compact Calabi-Yau three-fold $X_3$ outside of a compact submanifold of $Z_-$, and the Ricci-flat metric on $Z_-$ exponentially asymptotes to the Ricci-flat metric on the product manifold $I \times X_3$.} 
\\

Such manifolds were discussed in \cite{Kovalev2010}. Note that such $G_2$-manifolds are easily constructed as (resolutions of) orbifolds $(X_3 \times \R_t) /\Z_2$ with the $\Z_2$ acting as $t\rightarrow -t$ and as an anti-holomorphic involution on the Calabi-Yau three-fold $X_3$. The quotient by such involutions can be thought of as a Calabi-Yau three-fold $X_3$ fibered over an half-open interval which undergoes a degeneration at one end. In particular, it is not surprising if the Ricci-flat metric on such a resolution of $(X_3 \times \R)/ \Z_2$ asymptotes to the Ricci-flat metric on $X_3 \times \R$ far away from the origin of $\R$. A particularly simple case is given by freely acting anti-holomorphic involutions. In this case no resolution is required, and the Ricci flat metric on $(X_3 \times \R) /\Z_2$ is simply the quotient of the Ricci-flat metric on $X_3 \times \R$. The simplicity of this examples comes at a price, however, as the holonomy group of such acyl $G_2$-manifolds is not the full $G_2$ \cite{nordstrom_2008,Kovalev2010}, but only $SU(3) \rtimes Z_2$, so that we can call them acyl barely $G_2$-manifolds. Such acyl barely $G_2$-manifolds in particular have a nontrivial fundamental group $\Z_2$.
\\

We are now ready to propose our construction of what we will call generalized connected sum (GCS) $\Spin(7)$-manifolds. Take an acyl Calabi-Yau four-fold $Z_+$ with asymptotic neck region isomorphic to $I_{t_+} \times \bbS^1_{\theta_+} \times X_3^+$ for $t_+ = 0\cdots l$ and a acyl $G_2$-manifold $Z_-$ with asymptotic neck region isomorphic to $I_{t_-} \times X_3^-$ for $t_- = 0\cdots l$. Then $Z_+$ and $Z_- \times \bbS^1_{\theta_-}$ can be glued as topological manifolds to a manifold $Z$ by identifying the neck regions $I_{t_\pm} \times \bbS^1_{\theta_\pm} \times X_3^\pm$ such that 
\begin{equation}\label{eq:gluingZpm}
\begin{aligned}
t_+ & = l - t_- \\
\theta_+ & = \theta_- \\
\phi: \quad X_3^+ & \cong X_3^- 
\end{aligned}
\end{equation}
for a biholomorphic map $\phi$. If the Ricci-flat metrics on $Z^\pm$ asymptote to the Ricci-flat metrics on $X_3^\pm$, this takes us close to a Ricci-flat metric on $Z$. We conjecture that for long enough neck regions ($l$ large enough) there exists a Ricci-flat metric $g$ associated with a torsion free $\Spin(7)$ structure on $Z$, which is found by a small perturbation of the Ricci-flat metrics on $Z_+$ and $Z_-\times\bbS^1$. 

If $Z_\pm$ are both simply connected, it follows from the Seifert-van Kampen theorem  that $Z$ is a simply connected eight-manifold. This means that the holonomy group of $Z$ must be equal to $\Spin(7)$ (and not a subgroup) if $Z$ has a torsion-free $\Spin(7)$ structure.

Before exploring the consequences of our proposed construction further, let us briefly discuss the mathematical work needed to put our proposal on firm ground. The work \cite{nordstrom_2008,Kovalev2010,Haskins2015} on acyl Calabi-Yau four-folds and acyl $G_2$-manifolds and their deformation theory, together with clear criteria when the asymptotic Calabi-Yau three-folds $X_3^\pm$ allow a biholomorphic map $\phi$, should clarify under which circumstances a gluing can be found for a given pair of such manifolds. In our examples, we can easily find such diffeomorphisms by realizing $X_3^\pm$ as hypersurfaces in toric varieties, so that a diffeomorphic pair can be simply constructed by writing down identical algebraic equations. The crucial task is then to show the existence of a Ricci-flat metric with holonomy $\Spin(7)$ on the resulting topological manifold $Z$. By making the neck regions very long, we expect that the torsion introduced when gluing $Z_+$ and $Z_-\times\bbS^1$ can be made sufficiently small for a torsion-free $\Spin(7)$ structure to exist nearby.

As we shall discuss in more detail in section \ref{sec:CY34}, there are instances of GCS $\Spin(7)$-manifolds which can globally be realized as (resolutions) of quotients of Calabi-Yau four-folds. Quotients by free actions, where no resolution is needed, or situations in which resolutions with Ricci-flat metrics can be constructed provide non-trivial examples of our construction in which their existence is proven by other means.

\subsection{Calibrating Forms}

Both the acyl Calabi-Yau four-fold $Z_+$ and the product of the acyl $G_2$-manifold $Z_-$ with a circle have a torsion free $\Spin(7)$ structure. In this section we show how the identification \eqref{eq:gluingZpm} between the asymptotic neck regions produces a globally defined $\Spin(7)$ structure. 

Let us denote the K\"ahler form and holomorphic four-form of $Z_+$ by $\omega_+$ and $\Omega_+$. The Cayley four-form defining a torsion-free $\Spin(7)$ structure on $Z_+$ is then given by
\begin{equation}
\Psi_+ = \Re \Omega_+ + \tfrac12 \,\omega_+ \wedge \omega_+ \, .
\end{equation}
In the asymptotic neck region, we can introduce coordinates $t_+$ and $\theta_+$ and decompose the $SU(4)$ structure as
\begin{equation}
\ba
\omega_+ &= \omega_{3,+} + dt_+ \wedge d\theta_+  \\
\Omega_+ &= d\theta_+ \wedge\Re \Omega_{3,+} - dt_+ \wedge \Im \Omega_{3,+}   + i \left(  dt_+ \wedge \Re \Omega_{3,+}  + d\theta_+ \wedge \Im \Omega_{3,+}  \right)  \,,
\ea
\end{equation}
where $\omega_{3,+}$ and $\Omega_{3,+}$ are the K\"ahler form and holomorphic three-form on $X_3^+$. This means that the $\Spin(7)$ structure in the neck region is 
\begin{equation}\label{eq:psifromtheleft}
\Psi_+ =d\theta_+ \wedge\Re \Omega_{3,+} - dt_+ \wedge \Im \Omega_{3,+}  + \tfrac12 \,\omega_{3,+} \wedge \omega_{3,+} +  d\theta_+ \wedge dt_+  \wedge \omega_{3,+} \, .
\end{equation}

Similarly, the $\Spin(7)$ structure on $Z_- \times \bbS^1_{\theta_-}$ is given by 
\begin{equation}
\Psi_+ = d\theta_- \wedge \varphi_- + \ast \varphi_- 
\end{equation}
in terms of the $G_2$ structure $\varphi_-$ on $Z_-$. In the neck region, this is further decomposed as
\begin{equation}
\begin{aligned}
 \varphi_- &= -dt_- \wedge \omega_{3,-} + \Re \Omega_{3,-}  \\
 \ast\varphi_- &=  \tfrac12 \omega_{3,-}\wedge \omega_{3,-} + dt \wedge \Im \Omega_{3,-} 
\end{aligned}
\end{equation}
in terms of the K\"ahler form and holomorphic three-form on $X_{3}^-$. We hence find that the $\Spin(7)$ structure on the neck region of $Z_-\times \bbS^1$ can be written as
\begin{equation}\label{eq:psifromtheright}
\Psi_- = - d\theta_- \wedge dt_- \wedge \omega_{3,-} + d\theta_- \wedge \Re \Omega_{3,-} + \tfrac12 \omega_{3,-}\wedge \omega_{3,-} + dt \wedge \Im \Omega_{3,-} \, . 
\end{equation}
This means that the two $\Spin(7)$ structure $\Psi_\pm$ are consistently glued together under a diffeomorphism which identifies the neck regions as
\begin{equation}
\begin{aligned}
t_+ & = - t_- \\
\theta_+ & = \theta_- \\
\omega_{3,+} & = \omega_{3,-} \\
\Omega_{3,+} & = \Omega_{3,-}
\end{aligned}
\end{equation}
which is noting but the map \eqref{eq:gluingZpm} proposed earlier. 

\subsection{Topology of GCS $\Spin(7)$-Manifolds}

The easiest topological number to determine for the GCS $\Spin(7)$-manifold $Z$ is given by the Euler characteristic $\chi(Z)$. As it is additive, we have
\begin{equation}
\chi(Z)= \chi(Z_+)+\chi(Z_-\times \bbS^1) - \chi\left(Z_+ \cap (Z_- \times \bbS^1) \right) = \chi(Z_+) \, ,
\end{equation}
where we have used the fact that the Euler characteristic vanishes for any manifold with an $\bbS^1$ factor. 

Two copies $Z_+$, $Z_+'$ of the acyl Calabi-Yau four-fold $Z_+$ can be glued to a compact Calabi-Yau four-fold $X_4 = Z_+ \cup Z_+'$ such that $Z_+ \cap Z_+' = I \times \bbS^1 \times X_3$ for a Calabi-Yau three-fold $X_3$. It follows that $\chi(X_4) = 2 \chi(Z_+) = 2\chi(Z)$. As the Euler characteristic of any Calabi-Yau four-fold is divisible by six \cite{Klemm:1996ts}, this implies that $\chi(Z)$ is divisible by $3$, so that $b_4^-$ in  \eqref{eq:b4-intermsofchib2b3} is always an integer.

As we have defined an acyl Calabi-Yau four-fold to be simply connected, but left room for the possibility of an acyl $G_2$-manifold to have a non-trivial fundamental group, the Seifert-van Kampen theorem tells us that $Z$ is simply connected if an only if $Z_-$ is simply connected. If $Z_-$ is not simply connected, $Z$ can have a non-trivial fundamental group, which signals that the holonomy group of $Z$ is smaller than $\Spin(7)$ \cite{joyce1996spin7,joyce2000compact}. 

The easiest way to construct acyl $G_2$-manifolds is by a free quotient $X_3 \times \R$ by $\Z_2$, in which case the fundamental group of $Z_-$ is not trivial but equal to $\Z_2$. This works as follows. Consider two points identified by the anti-holomorphic involution on $X_3$ over the origin of $\R$. Any path connecting two such points becomes a closed loop in the quotient and, as the involution acts freely, cannot be homotopic to a point. If $X_3$ is simply connected, all such loops are homotopic, so that $\pi_1((X_3 \times \R)/\Z_2) = \Z_2$, which implies that $Z_-$ only has holonomy group $SU(3) \rtimes \Z_2$, i.e. is a barely $G_2$ manifold. On $Z$, such loops give rise to a non-trivial fundamental group, which is $\Z_2$ as well, so that $Z$ does not have the full holonomy group $\Spin(7)$. From the point of view of physics, compactification on such `barely' $\Spin(7)$-manifolds does not give rise to extended supersymmetry as there is still only one covariantly constant spinor. In fact, it is not hard to see that $Z$ has holonomy group $SU(4) \rtimes Z_2$ in this case: the holonomy group $SU(4)$ of $Z_+$ and the holonomy group $SU(3) \rtimes \Z_2$ of $Z_-$ share a common $SU(3)$, the holonomy group of $X_3$. 

We can compute the cohomology groups of $Z$ in terms of the cohomology groups of $Z_\pm$ and the pull-backs to $X_3$ by using the Mayer-Vietoris exact sequence, which implies that
\begin{equation}\label{eq:cohomZfromgammas}
H^i(Z,\Z) = \ker \gamma^i \oplus \coker \gamma^{i-1} \, ,
\end{equation}
where 
\begin{equation}
\gamma^i :\,\, H^i(Z_+,\Z) \oplus H^i(Z_-\times\bbS^1,\Z) \rightarrow H^i(X_3 \times \bbS^1 ,\Z) 
\end{equation}
can be expressed in terms of the restriction maps 
\begin{equation}
\begin{aligned}
\beta_+^i : &&H^i(Z_+,\Z) &\rightarrow H^i(X_3 \times \mathbb{S}^1,\Z)  \\
\beta_-^i : &&H^i(Z_-,\Z) &\rightarrow H^i(X_3,\Z)  \,.
\end{aligned}
\end{equation}
Note that both $\beta_-^i$ and $\beta_-^{i-1}$ feature in $\gamma^i$ due to the product $\bbS^1$. 

Under a few assumptions, which are met in the examples to be discussed later, we can now explicitly work out the various contributions to $H^\bullet(Z)$. First of all, let us assume that the acyl Calabi-Yau four-fold $Z_+$ can be constructed from an algebraic four-fold $\tilde{Z}$ which is fibered by Calabi-Yau three-folds by excising a fiber $X_3$. Furthermore, let us assume that the images of $\beta_+^2$ and $\beta_+^4$ are surjective and that $b^3(Z_+)=b^5(Z_+) = 0$. 

As noted above, we assume that we can construct $Z_-$ as (a resolution) of the quotient $( X_3 \times \R_t )/ \Z_2$ by an anti-holomorphic involution. The resolution must be such that it preserves a metric of holonomy $G_2$ on $Z_-$ and the property of $Z_-$ being asymptotically cylindrical. We will denote the numbers of even/odd classes of $X_3$ under this $\Z_2$ by $b^i_e$ and $b^i_o$, respectively. Finally, the kernels of the restriction maps $\beta_\pm^i$ and their ranks are abbreviated as
\begin{equation}
 N_\pm^i  \equiv \ker \beta_\pm^i\, \hspace{1cm} n_\pm^i \equiv |N_\pm^i|\, .
\end{equation}
With this notation and under the assumptions we have made, \eqref{eq:cohomZfromgammas} implies that
\begin{equation}\label{eq:bettisgcsspin7}
\begin{aligned}
b^1(Z) & = 0 \\  
b^2(Z) & = n_+^2 + n_-^2 + b^2_e \\  
b^3(Z) & = n_-^2 + n_-^3\\  
b^4(Z) & = n_-^3 + n_-^4 + n_+^4 + b^2_o + b^3_o + b^3_e + b^4_e\\  
b^5(Z) & = n_-^4 + n_-^5 \\  
b^6(Z) & = n_+^6 +n_-^5 + b^4_o \\  
b^7(Z) & = 0 \,.
\end{aligned}
\end{equation}
This can be seen as follows. For $H^1(Z)$, the unique class in $H^1(X_3 \times\bbS^1)$ is in the image of $\gamma^1$ and solely originates from $H^1(Z_-\times\bbS^1$). Hence $|\ker(\gamma^1)| = 0$, which together with $\coker(\gamma^0)=0$ implies $b^1(Z)=0$. 

Let us now consider $b^2(Z)$. As the unique class in $H^1(Z_-\times\bbS^1)$ is in the image of $\gamma^1$, we find that $\coker(\gamma^1)=0$ so that $H^2(Z) = \ker(\gamma^2)$ follows. In turn, $\ker(\gamma^2)$ has three different contributions: 
\begin{equation}
\ker(\gamma^2) = \ker (\beta_+^2) \oplus  \ker (\beta_-^2)  \oplus \left[ \im (\beta_+^2) \cap \im (\beta_-^2) \right] \, .
\end{equation}
As $\beta_+^2$ is surjective by assumption, the last term is simply given $b^2_e$ and we recover the expression in \eqref{eq:bettisgcsspin7}. 

The computation for $b^3(Z)$ is made particularly simple by our assumptions. As $|\coker(\gamma^2)| = 0$, it only receives a contribution from $\ker(\gamma^3)$. As $b^3(Z_+)=0$, the only contribution to $b^3(Z)$ is from $\ker(\beta^2_-)\oplus\ker(\beta^3_-) = N_-^2 \oplus N_-^3$. 

For $b^4(Z)$, all potential contributions are non-trivial. First of all, there is a contribution $n_+^4$ from $\ker \beta_+^4$, as well as $n_-^3 + n_-^4$ from $\ker \beta_-^3  + \ker \beta_-^4 $. Furthermore, there are $b^2_o$ classes in $\coker(\gamma^3)$, which correspond to three-forms with one leg on the product $\bbS^1$ of the cylinder region, and $b^3_o$ classes in $\coker(\gamma^3)$ which correspond to three-forms purely on $X_3$. Finally, there is a contribution
\begin{equation}
 (\im \beta_+^4) \cap \left[ (\im \beta_-^4) \oplus (\im \beta_-^3) \right] \, ,
\end{equation}
which simply has dimension $b^4_e + b^3_e$ as $\beta_+^4$ is surjective. Together, all of these contributions give the expression for $b^4(Z)$ in \eqref{eq:bettisgcsspin7}.

The computation of $b^5(Z)$ again only receives a contribution from $\ker \gamma^5$ as $\coker \gamma^4$ is empty by assumption. Furthermore, $\gamma^5$ is only non-trivial on $Z_- \otimes \bbS^1$, so that $|\ker \gamma^5| = n_-^4 + n_-^5$.

The computation of $b^6(Z)$ is similar to $b^2(Z)$, it receives a term $b^4_o$ from $\coker(\gamma^5)$, as well as terms $n_-^5$ and $n^6_+$ from $\ker(\gamma^6)$. Note that $N^6_- = 0$ as $Z_-$ is a $G_2$ manifold. 

Finally, $\coker(\gamma^6) = 0$ and $\ker(\gamma^7)=0$ gives $b^7(Z)=0$. 
\\
\\
The various contributions $n_-^i$ and $h^i_e,h^i_o$ are not all independent and their relations make Poincar\'e duality for \eqref{eq:bettisgcsspin7} manifest. For anti-holomorphic involutions of a Calabi-Yau three-fold $X_3$, the Lefschetz fixed-point theorem combined with Poincar\'e duality implies that $b^2_o = b^4_e$ and $b^2_e = b^4_o$. Furthermore, Poincar\'e duality on the compact $G_2$ manifold $(X_3 \times \bbS^1)/\Z_2$ implies that $n_-^2 = n_-^5$ and $n_-^3 = n_-^4$. Together with $n_+^2 = n^6_+$, these relations imply that $b^i(Z) = b^{8-i}(Z)$ in \eqref{eq:bettisgcsspin7}.

The number of deformations of the Ricci-flat metric is given by $b^4_-+1$ for manifolds of $\Spin(7)$-holonomy. Using \eqref{eq:b4-intermsofchib2b3}, this can be written as
\begin{equation}
b^4_- +1 = -8 + \tfrac13 \left(2-b^2+b^3+b^4 \right) \, ,
\end{equation}
which becomes
\begin{equation}\label{eq:b4-intermsofbbdata}
b^4_- +1 = -8 + n^3_- + \tfrac13 \left(2+n^4_+ - n^2_+ + 2 b^2_o-b^2_e + b^3_e + b^3_o \right)
\end{equation}
in our GCS-construction. The Euler characteristic of $Z$ is given by
\begin{equation}
\chi(Z) = 2 + 2 b^2(X_3) + b^3(X_3) + n^4_+ + 2 n^2_+ \, ,
\end{equation}
and only depends on the data of $Z_+$, as expected.

{Even though the topology of the resolutions of Joyce's $\Spin(7)$-orbifolds discussed in section \ref{sec:JoyceExample} is most conveniently computed otherwise, one can also use their decomposition as GCS $\Spin(7)$-manifolds together with the relations of this sections to find their topology from resolutions of two building blocks $Z_\pm$.}

\subsection{3d Field Theory and Sectors of Enhanced Supersymmetry}
\label{sec:FieldTheory}

In compactifications of M-Theory on a smooth $\Spin(7)$-manifold, the low energy effective theory at the classical level gives a 3d $\mathcal{N}=1$ supergravity theory with $n_v = b^2(Z)$ massless $U(1)$ vector multiplets and $n_r = b^3(Z)+b^4_-(Z)+1$ massless uncharged real scalar multiplets. As the $\Spin(7)$-manifolds considered here are glued from pieces with holonomy $SU(4)$ and $G_2$, respectively, we expect to find subsectors of enhanced supersymmetry in our spectrum, similar to the observations made in \cite{Guio:2017zfn} regarding TCS $G_2$ manifolds, to arise from localized forms in the building blocks.

Consider first the multiplets arising from localized forms on $Z_- \times \bbS^1$. By \eqref{eq:bettisgcsspin7}, each two-form in $N^2_-$ gives rise to both a two-form and a three-form on $Z$, which combine to form the bosonic field content of a 3d $\mathcal{N}=2$ vector multiplet. Furthermore, each three-form in $N^3_-$ gives rise to a real scalar due to its appearance in the formula for $b^3(Z)$ in \eqref{eq:bettisgcsspin7}. As for compactifications of M-Theory on a $G_2$-manifold times a circle, the moduli associated to these three-forms pair up with deformations of the metric to form 3d $\mathcal{N}=2$ chiral multiplets: this can be seen from \eqref{eq:b4-intermsofbbdata}, which shows that each three-form in $N^3_-$ corresponds to an anti-self-dual four-form, i.e. a deformation of the metric. 

Let us now turn to multiplets which arise from localized forms on $Z_+$. Each two-form of $Z_+$ in $N_+^2$ gives rise to a 3d $\mathcal{N}=1$ vector multiplet due to its contribution to $b^2(Z)$. As for compactifications of M-Theory on Calabi-Yau four-folds, we expect this degree of freedom to pair up with a real scalar to the bosonic degrees of freedom of a 3d $\mathcal{N}=2$ vector multiplet. This degree of freedom must originate from an anti-self-dual four-form in $N^4_+$. Furthermore, we expect the remaining anti-self-dual four-forms in $N^4_+$ to appear pairwise, so as to combine into
3d $\mathcal{N}=2$ chiral multiplets. 

To see how this comes about, we need to exploit the fact that $Z_+$ and its compactification $\tilde{Z}_+$ are K\"ahler manifolds and carry a Lefschetz $SU(2)$ action. The upshot is that the anti-self-dual four-forms on $\tilde{Z}_+$ are given by the four-forms in $H^{3,1}(\tilde{Z}_+)\oplus H^{1,3}(\tilde{Z}_+)$, together with four-forms of the type $\omega_k \wedge J$, where $J$ is the K\"ahler form on $\tilde{Z}_+$ and the $\omega_k$ are $(1,1)$ forms on $\tilde{Z}_+$ such that $\omega_k \wedge J^3=0$. As every two-form in $N_+^2$ gives a four-form in $N_+^4$ upon wedging with $J$, we can hence associate an anti-self-dual four-form in $N_+^4$ on $Z_+$ to every element of $N_+^2$ (modulo the $(1,1)$ forms in the image of $\beta^2_+$). The remaining anti-self-dual four-forms in $N_+^4$ then have to appear pairwise, as they must correspond to $H^{3,1}(\tilde{Z}_+)\oplus H^{1,3}(\tilde{Z}_+)$. As we can think about the forms in $N^4_+$ as being localized on the acyl Calabi-Yau four-fold $Z_+$ far away from the gluing region, (anti) self-duality on $Z_+$ will imply (anti) self-duality of the corresponding forms on $Z$. 

Note that this argument precisely reflects how these degrees of freedom originate in Physics. For two-forms in $N_+^2$ and self-dual four-forms in $N_+^4$, there are deformations of the Ricci-flat metric of $Z_+$ (K\"ahler and complex structure deformations), which do not alter the cylindrical region in which $Z_+$ is glued to $Z_-\times \bbS^1$. As we expect the Ricci-flat metric of $Z$ to be well approximated by the Ricci-flat metrics of $Z_+$ far away from the gluing region, these become deformations of the $\Spin(7)$-manifold $Z$ if we stretch the neck region sufficiently long. The metric deformations associated with $N_+^2$ and $N_+^4$, together with the 3d vectors originating from the $C$-field on $N_+^2$, only see the geometry of a Calabi-Yau four-fold and hence give rise to a subsector with 3d $\mathcal{N}=2$ supersymmetry. As the property of forms being (anti) self-dual is a topological constraint, the counting and identifications we have performed will persists throughout the moduli space of $Z$, so that the subsectors with enhanced $\mathcal{N}=2$ supersymmetry in 3d will persist. We have summarized the result of this discussion in table \ref{tab:N2N4}

\begin{table}\begin{center}
\begin{tabular}{c|c}
3d Multiplets & Origin \\
\hline
$\mathcal{N}=2$  vector & $N^2_+$ \\
$\mathcal{N}=2$ chiral & $h^{3,1}{(\tilde{Z}_+)}$ \\\hline
$\mathcal{N}=4$  vector & {$b_2(X_3)$, $b_3 (X_3)$} \\\hline
$\mathcal{N}=2$  vector & $N^2_-$ \\
$\mathcal{N}=2$ chiral & $N^3_-$ \\
\end{tabular}
\caption{Subsectors with enhanced supersymmetry and their topological origin. 
The $\mathcal{N}=4$ vector multiple from the neck region is only massless in the infinite neck limit and away from this becomes an $\mathcal{N}=1$ vector. \label{tab:N2N4}}
\end{center}
\end{table}

The remaining moduli of M-Theory on the GCS $\Spin(7)$-manifold $Z$ do not appear in multiplets with enhanced supersymmetry, as they are associated with the gluing along $X_3 \times \bbS^1$. These moduli are, however, in one-to-one correspondence with the geometric deformations of the Calabi-Yau three-fold $X_3$ in the neck region, i.e. K\"ahler and complex structure deformations of $X_3$. While compactification of M-Theory on $X_3 \times \bbS^1 \times \bbS^1$ results in a theory with $\mathcal{N}=4$ supersymmetry in 3d, these multiplets are truncated to $\mathcal{N}=1$ for M-Theory compactified on $Z$. Similar to the case of TCS $G_2$ manifolds \cite{Guio:2017zfn}, the eigenvalues under the Laplace operator of the appropriate non-harmonic forms needed to lift these $\mathcal{N}=1$ multiplets to $\mathcal{N}=4$ multiplets go to zero in the limit in which the neck region is stretched.  

\subsection{GCS $\Spin(7)$-Manifolds as Quotients of $\CY_4$}\label{sect:gcsasquotients}
\label{sec:CY34}

Our construction has a natural connection to $\Spin(7)$-manifolds constructed as quotients of Calabi-Yau four-folds by anti-holomorphic involutions, however we will see that there are key difference to the previous constructions in \cite{joyce1996spin7_new,Bonetti:2013fma}. Consider a compact Calabi-Yau four-fold $X_4$, which is fibered by Calabi-Yau three-folds over a base $\P^1$ with homogeneous coordinates $[z_1:z_2]$ and let $z=z_1/z_2$. Let us denote the fiber over a point $p$ of the $\P^1$ base by $X_3(p)$. For appropriate fibrations, we may then tune the complex structure (defining equation) of $X_4$ such that 
\begin{equation}
X_3(z) = X_3(1/\bar{z}) \, .  
\end{equation}
For appropriate anti-holomorphic involutions $\sigma$ acting on $X_3(p)$ for all $p$, we may then construct an anti-holomorphic involution 
\begin{equation}
\Sigma = \sigma \circ \left\{ \begin{aligned}
                      z_1 \rightarrow \bar{z}_2 \\
		      z_2 \rightarrow \bar{z}_1 
                     \end{aligned}
\right\}
\end{equation}
acting in $X_4$. This will produce a singular $\Spin(7)$-manifold $Z_s = X_4 / \Sigma$, and we will assume that it can be resolved into a smooth $\Spin(7)$-manifold $Z$. 

In the $\P^1$ base the fixed locus of this involution will be given by the circle $\bbS_f^1 = \{|z|^2 = 1\}$ and the quotient effectively truncates the base $\P^1$ to a disc with boundary. We may make this disc very large and furthermore confine all of the singular fibers of the fibration of $X_3(p)$ near the origin. If the fibration of $X_3(p)$ over the fixed $\bbS^1_f$ is trivial, the fibers over $\bbS^1_f$ all become identical and we may simply denote them by $X_3$. Cutting along a circle of fixed radius now produces one half near the boundary of the disc which may be described as 
\begin{equation}
\bbS^1_f \times (X_3 \times \R)/ \sigma \, .
\end{equation}
In the limit in which the disc is very large, a resolution of $Z_s$ to $Z$ is equivalent to a resolution of $(X_3 \times \R)/ \sigma$ to a smooth acyl $G_2$ manifold $Z_-$. The other half of $Z$ near the origin of the disc does not require resolution and becomes an acyl Calabi-Yau four-fold $Z_+$ which asymptotes to $X_3$. In the middle region, the two spaces $Z_+$ and $Z_-\times\bbS^1_f$ are glued along the product of an interval and 
$X_3 \times \bbS^1_f$. This is precisely the gluing construction we have given above. An illustration is given in figure \ref{fig:CY4Pic}.

Let us highlight two aspects of the topology of $Z$ which are immediately recovered from this point of view. The resolution of $Z_s$ to $Z$ only introduces new cycles sitting over a circle, so that the topological Euler characteristic of $Z$ is the same as that of $Z_s$. But the topological Euler characteristic of $Z_s$ is simply given by half of the Euler characteristic of $X_4$. In fact, we may think of $X_4$ as being glued from two copies of the acyl Calabi-Yau four-fold $Z_+$, which implies that the Euler characteristic of $Z$ is equal to that of $Z_+$, which is the same result found from our gluing construction earlier. Furthermore, we have observed that we produce a barely $\Spin(7)$-manifold with fundamental group $\Z_2$, whenever $Z_-$ is constructed from a free quotient of $X_3 \times \R$ by $\sigma$. This means that we may write $Z$ as the free quotient $X_4 /\Sigma$, so that we immediately recover $\pi_1(Z) =\Z_2$.

As we have explained, a GCS $\Spin(7)$-manifold can be described as a resolution of a quotient of a Calabi-Yau four-fold by an anti-holomorphic involution whenever the acyl $G_2$ manifold $Z_-$ can be described as a resolution of $(X_3 \times \R)/ \sigma$ for an anti-holomorphic involution $\sigma$. In the absence of a different construction for acyl $G_2$ manifolds, our GCS construction is hence equivalent to forming quotients of a specific class of Calabi-Yau four-folds. In this context, it implies that a resolution of singularities of the quotient by $\Sigma$ is already captured by resolving quotients $(X_3 \times \R)/ \sigma$. Furthermore, it gives a completely new point of view on such geometries; it shows how to distinguish subsectors of enhanced supersymmetry and, as we shall see, allows to construct dual heterotic backgrounds on TCS $G_2$ manifolds. Finally, the realization that many GCS $\Spin(7)$-manifolds are simply (resolutions of) quotients $X_4 /\Sigma$ in fact proves that our construction produces eight-dimensional manifolds with Ricci-flat metrics with holonomy group $\Spin(7)$ (or $SU(4) \rtimes Z_2$).

In \cite{joyce1996spin7_new} Joyce proposed another construction of $\Spin(7)$-manifolds base on non-free quotients of Calabi-Yau four-folds with orbifold singularities was given. Crucially, the resulting singular $\Spin(7)$ orbifolds are taken to only have isolated singularities here, which can be resolved by gluing in appropriate ALE spaces. Even though similar in spirit, the anti-holomorphic involution appearing here are either free or have fixed loci of real dimension at least one, so these constructions are distinct. It seems likely, however, that there are examples for which the resolutions can be described by gluing in a acyl $G_2$ manifold times a circle, so that there will be many $\Spin(7)$-manifolds which can be found using both constructions. 

\subsection{A Simple Example}\label{sec:NewSpin7}

We now construct a new $\Spin(7)$-holonomy manifold using the GCS construction. There will be further examples in the next section, which are geared toward the application of M-theory/Heterotic duality. 

Consider a smooth anti-canonical hypersurface in weighted projective space $\mathbb{P}_{11114}$. The resulting space is a Calabi-Yau three-fold $X_{1,149}$ with Hodge numbers $h^{1,1}=1$ and $h^{2,1}=149$. We now show how to construct an acyl $G_2$-manifold $Z_-$ which asymptotes to $X_{1,149}\times \R$ and an acyl Calabi-Yau four-fold $Z_+$ which asymptotes to $X_{1,149} \times \bbS^1 \times \R$, and then glue $Z_+$ and $Z_-\times\bbS^1$ to a $\Spin(7)$-manifold $Z$ as described above. Let us first describe the acyl $G_2$-manifold $Z_-$. The Calabi-Yau three-fold $X_{1,149}$ admits a freely acting anti-holomorphic involution, which is obvious by choosing a Fermat type hypersurface
\begin{equation}\label{eq:x1149_sym}
x_1^{8}  +  x_2^{8} + x_3^{8} + x_4^{8} + x_5^{2} = 0\,,
\end{equation}
where $[x_1:x_2:x_3:x_4:x_5]$ are the homogeneous coordinates of $\mathbb{P}_{11114}$. Letting $x_i \rightarrow \bar{x}_i$ then acts as a fixed-point free anti-holomorphic involution, as the above equation has no solutions purely over the reals. The odd/even Hodge numbers are 
\be
h^2_e = 0 \,,\qquad h_o^2 = 1 \,,\qquad 
h^3_e =  h_o^3 = 149 \, ,
\ee
and $(X_{1,149} \times \R)/\Z_2$ is an acyl $G_2$-manifold. Note that $n_-^i = 0$ for all $i$ as no resolution of singularities is required on $Z_-$.

To find an acyl Calabi-Yau four-fold, we consider an algebraic four-fold $\tilde{Z}_+$ which is a fibration of $X_{1,149}$ over $\P^1$ and has $c_1(\tilde{Z}_+) = [X_{1,149}]$. Such a $\tilde{Z}_+$ can easily be constructed by as the zero locus of a generic section of $\mathcal{O}(8,1)$ over $\mathbb{P}_{11114} \times \P^1$. Such a four-fold has $h^{1,1}(\tilde{Z}_+)=3$, which means that $n_+^2=1$. Furthermore, the resulting acyl Calabi-Yau four-fold $Z_+ = \tilde{Z}_+\setminus X_{1,149}$ has $\chi(Z_+)= \chi(\tilde{Z}_+) - \chi(X_{1,149}) = 1680 = \chi(Z)$. 

Note that we may choose an element of the algebraic family $\tilde{Z}_+$ such that the fiber over some chosen point in the base $\P^1$ is described by \eqref{eq:x1149_sym}, so that $Z_+$ manifestly asymptotes to a Calabi-Yau three-fold which is isomorphic to the Calabi-Yau three-fold in the neck region of $Z_-$. For the resulting GCS $\Spin(7)$-manifold constructed from this choice of $Z_+$ and $Z_-$, it now follows that 
\begin{equation}\label{eq:nondualex1betti}
\begin{aligned}
b^2(Z) = 1 \hspace{.5cm}  b^3(Z) = 0 \hspace{.5cm}  b^4_-(Z) = 551  \hspace{.5cm}  b^4_+(Z) = 1125 \, .
\end{aligned}
\end{equation}
Using \eqref{eq:b4-intermsofchib2b3}, this is most efficiently calculated by using $\chi(Z) = 1680$. One can also work out that $n_+^4 = 1374$, which reproduces $b^4(Z) = 1676$ using \eqref{eq:bettisgcsspin7}. Due to the free action of the anti-holomorphic involution on $X_{1,149}$, this example produces a barely $\Spin(7)$-manifold. By considering more general anti-holomorphic involutions and resolving the resulting singularities, similar examples with full holonomy $\Spin(7)$ can be constructed from this procedure.

In agreement with our general discussion in section \ref{sect:gcsasquotients}, $Z$ may also be found as the \emph{free} quotient of a Calabi-Yau four-fold $X_4$ realized as a hypersurface in $\mathbb{P}_{11114} \times \P^1$. The Hodge numbers of $X_4$ are
\begin{equation}
h^{1,1}(X_4) = 4 \hspace{.5cm} h^{2,1}(X_4) = 0 \hspace{.5cm} h^{3,1}(X_4) = 548 \hspace{.5cm}  h^{2,2}(X_4) = 2252
\end{equation}
and $\chi(X_4) = 3360$. The free involution $\Sigma$ acts as 
\begin{equation}
\begin{aligned}
x_i &\rightarrow \bar{x}_i \\
z_1 &\rightarrow \bar{z}_2 \\
z_2 &\rightarrow \bar{z}_1 \, ,
\end{aligned}
\end{equation}
where $[z_1:z_2]$ are homogeneous coordinates on the base $\P^1$ and $x_i$ are the homogeneous coordinates on $\mathbb{P}_{11114}$. We may choose a smooth invariant hypersurface on which $\Sigma$ acts without fixed points as
\begin{equation}
(z_1^2 + a_5 z_1 z_2 + z_2^2) x_5^{2} + \sum_i^4 (z_1^2 + a_i z_1 z_2 + z_2^2) x_i^{8}  = 0 \, ,
\end{equation}
for a set of real parameters $a_i$. A little thought reveals that a one-dimensional subspace of the forms in $H^{1,1}(X_4)$ is even under $\Sigma$, which together with $\chi(X_4) = 3360$ and $b^3(X_4)=0$ implies \eqref{eq:nondualex1betti}.


\section{M-theory on GCS $\Spin(7)$/Heterotic on TCS $G_2$}
\label{sec:MHet}

In this section we construct a new class of GCS $\Spin(7)$ manifolds $Z$ following our general discussion in section \ref{sect:generalconstruction}, where the building blocks are an open acyl Calabi-Yau four-fold $Z_+$ and an open acyl $G_2$-manifold $Z_-$ (times a circle), which are both asymptote to a cylinder times the same Calabi-Yau three-fold $X_3$. The additional input in this section is that we will consider these in the context of the duality to Heterotic strings theory on $G_2$-manifolds, which we show to be TCS-manifolds, with building blocks $X_\pm$. 

Our strategy in finding dual compactifications will be similar to the one we used in \cite{Braun:2017uku}, i.e. using fiberwise duality between M-theory and heterotic string theory, we will identify dual pairs of building blocks, which can then be glued to find the dual $\Spin(7)$ and $G_2$-manifolds. As this requires to carefully dissect geometries on both sides, the following discussion will unfortunately be rather technical. After motivating our construction for a dual pair of geometries, we will summarize our results in Section \ref{sect:example1} before checking the spectra. 
In our first example, the $E_8 \times E_8$ gauge symmetry is broken by a (quotient of a) Wilson line, and we can easily match the light degrees of freedom of M-theory to those of the dual heterotic string theory. In general, such a match is much more difficult, as it requires to determine the number of moduli associated with a vector bundle on a $G_2$-manifold. We present a second closely related example in which we use the proposed duality to instead predict the number of bundle moduli on the $G_2$ manifold. 

We begin with a brief summary of heterotic $G_2$-systems in section \ref{sec:HetG2} and a discussion of the duality in section \ref{sec:MHetDual}, before then constructing examples of dual pairs.


\subsection{Heterotic String Theory on $G_2$-Manifolds}
\label{sec:HetG2}

Before we begin with a detailed discussion of dual pairs, we should explain how to construct   3d $\mathcal{N}=1$ theories from heterotic string theory on a $G_2$-holonomy manifold together with a vector bundle. Here we will discuss briefly  heterotic compactifications on $G_2$-holonomy manifolds. 

A compact orientable 7-dimensional manifold $J$ which has a Ricci-flat metric $g$ with holonomy group contained in $G_2$ supports a closed three-form $\Phi$, the Hodge-$\ast$-dual of which is closed as well, which can be expressed in local coordinates (in which $g$ is the euclidean metric) as
\begin{equation}
\begin{aligned}
\Phi = dx_{123}+dx_{145}+dx_{167}+dx_{246}-dx_{257}-dx_{347}-dx_{356}\, .
\end{aligned}
\end{equation}
This three-form and its Hodge-$\ast$-dual are preserved by the action of $G_2$. Besides having a Ricci-flat metric, $G_2$-manifolds support a single covariantly constant spinor. 

In contrast to $\Spin(7)$-manifolds, a necessary and sufficient condition for the holonomy group of a manifold $J$ with a Ricci-flat metric $g$ and $\hbox{hol}(g) \subseteq G_2$,to be equal to $G_2$ is that the fundamental group of $J$ is finite rather than trivial. We will refer to manifolds which support a single covariantly constant spinor, but have holonomy group $SU(3) \rtimes Z_2$ as barely $G_2$-manifolds. The only non-trivial Betti numbers of a $G_2$-manifold are $b^2$ and $b^3$ as $b^1 =0$, and the moduli space of Ricci-flat metrics for a $G_2$-manifold has real dimension $b_3$. 
For $G_2$-manifolds, calibrated submanifolds can have dimension three (associative cycles), in which case they are calibrated by $\Phi$, oder dimension four (co-associative cycles), in which case they are calibrated by $\ast \Phi$. 

Let us now discuss compactification of heterotic $E_8 \times E_8$ strings on $G_2$-manifolds. 
Earlier studies of this in the context of resolutions of Joyce orbifolds $T^7/\Gamma$ have appeared in \cite{Acharya:1996ef, Font:2010sj, Gemmer:2013ica}. The most in depth analysis of the conditions that such a compactification needs to satisfy were derived in \cite{delaOssa:2017pqy} and are given in terms of a {\it heterotic $G_2$-system}. This is comprised of a $G_2$-holonomy manifold and its tangent bundle $(J, TJ)$ as well as an $E_8\times E_8$-vector bundle with connection $(V, A)$. The curvatures $R$ and $F$ of $TJ$ and $V$ satisfy 
\be
R\wedge \ast \Phi =0 \,,\qquad F\wedge \ast \Phi =0 \,.
\ee
Furthermore there can be  NS5-branes wrapped on associative three-cycles, subject to the anomaly condition\begin{equation}
{\alpha' \over 4} \left( \Tr F \wedge F -  \Tr R \wedge R \right)= dH  \,,
\end{equation}
where $dH=$[NS5] the Poincar\'e dual of the homology class corresponding to associatives wrapped by NS5-branes.


\subsection{GCS $\Spin(7)$-Manifolds and M-theory/Heterotic Duality}
\label{sec:MHetDual}

Up to this point, our construction of GCS $\Spin(7)$-manifold was largely motivated for constructions of new geometries. An alternative motivation is to realize that these give rise to M-theory duals of heterotic strings on TCS $G_2$-manifolds. A review of the TCS construction is given in appendix \ref{app:TCS}.
Let us hence consider heterotic strings compactified on a manifold of $G_2$ holonomy $J$, together with an appropriately chosen vector bundle, giving an effective 3d  theory with $\mathcal{N} =1$ supersymmetry. Fiberwise duality to M-theory is possible if $J$ admits a calibrated fibration by three-tori $T^3$, which are replaced by K3 surfaces in the dual compactification of M-theory. As 3d compactifications of M-theory with $\mathcal{N} =1$ supersymmetry are found upon compactification on manifolds of $\Spin(7)$ holonomy, we expect the resulting geometry to be in this class. 

We already discussed the 7d duality between M/K3 and Het/$T^3$ in section \ref{sec:Setup}. We now apply this fiberwise to construct dual pairs of TCS and GCS constructions, in cases when there is a K3-fibration on the GCS $\Spin(7)$-manifold and dually on the heterotic side there is a $T^3$-fibered TCS $G_2$-manifold. In this section we make some initial observations about the general features of these compactifications. 

If $J$ is a TCS $G_2$-manifold, mirror symmetry for Type II strings implies the existence of a calibrated $T^3$-fibration, which realizes a mirror map in the spirit of Strominger--Yau--Zaslow (SYZ) \cite{Strominger:1996it} via three T-dualities \cite{Braun:2017csz}. Such a fibration exists if one of the two building blocks, say  $X_+$, used in the TCS-construction of $J$ is fibered by K3 surfaces $S_+$, which in turn admit an  fibration by an elliptic curve $\bbE$
\be
\mathbb{E} \hookrightarrow S_+ \rightarrow \mathbb{P}^1  \,.
\ee
This implies that $X_+$ admits an elliptic fibration with fiber $\bbE$ as well, and the $T^3$-fibers of $J$ restricted to $\bbE \times \bbS^1_{e,+}$ on $X_+ \times \bbS^1_{e,+}$. The Donaldson matching \eqref{eq:Dmatching} then maps $\bbE$ to a sLag fiber of the K3-surface $S_-$, and $\bbS^1_{e,+}$ to $\bbS^1_{b,-}$. Going away from the neck region of $X_-$, $\bbE \times \bbS^1_{b,-}$ becomes the SYZ fiber of $X_-$. 

Next let us consider how this maps under Heterotic/M-theory duality: 
\begin{itemize}
\item $X_+$: we replace the $T^3$-fibration given by $\bbE \times \bbS^1_{e,+}$ on $X_+ \times \bbS^1_{e,+}$, with a K3 surface $S_+$. Then  only two out of the three forms of the hyper-K\"ahler structure of $S_+$ have a non-trivial fibration over the base. 
In particular, we may choose to described the resulting K3 fibration algebraically and end up replacing the product of the acyl Calabi-Yau three-fold $X_+$ with $\bbS^1_{e,+}$ by an acyl Calabi-Yau four-fold $Z_+$. 
\item $X_-$: here, the SYZ-fiber of $X_-$ is fibered such that application of Heterotic/M-theory duality leads to a K3-fibration in which all three (1,1)-forms of the 
hyper-K\"ahler structure have a non-trivial variation over the base. As in the 4d $\mathcal{N}=1$ duality between heterotic strings and M-theory, we replace a Calabi-Yau three-fold by a $G_2$-manifold. In the present setup, this means replacing an acyl Calabi-Yau three-fold $X_-$ with an acyl $G_2$-manifold $Z_-$. In the cylinder region of $J$, where $J$ is diffeomorphic to a product $S \times \bbS^1 \times \bbS^1 \times I$ for a K3-surface $S$ with a calibrated $T^2$-fibration\footnote{Depending on a choice of complex structure, this is an elliptic or sLag fibration.}, we simply find $X_3 \times \bbS^1 \times I$ by application of Heterotic/M-theory duality. 
\end{itemize}
We hence find the statement that M-theory duals of heterotic strings on TCS $G_2$-manifolds are compactified on $\Spin(7)$-manifolds, which allows a construction as proposed in our GCS-construction: 
an acyl Calabi-Yau four-fold $Z_+$ and an acyl $G_2$-manifold $Z_-$ times an $\bbS^1$ are glued along a cylinder times a Calabi-Yau three-fold. 

In our discussion, we have so far ignored that the heterotic compactification comes equipped with a vector bundle on $J$, and we need to identify what this is mapped to on the dual M-theory geometry. For $X_+$, where only $\bbE$ varies non-trivially, replacing $X_+ \times \bbS^1_{e,+}$ by $Z_+$ is simply a non-compact version of the 3d duality between heterotic strings and M-theory on Calabi-Yau four-folds. 
We can hence apply the usual logic of how bundles constructed from spectral covers are translated in F-theory \cite{Friedman:1997yq}. Restricting $J$ to $X_- \times \bbS^1_{e,-}$, holomorphic vector bundles on a (non-compact) Calabi-Yau three-fold get translated to the geometry of a (non-compact) $G_2$-manifold, similar to discussed in \cite{Braun:2017uku}. Gluing $Z_+$ and $Z_- \times\bbS^1$ to a compact $\Spin(7)$-manifold $Z$ then implies that the geometry of $Z$ determines a {heterotic $G_2$-system} on the $G_2$-manifold $J$. It is hence tempting to construct such vector bundles on twisted connected sum $G_2$-manifolds by appropriately gluing holomorphic vector bundles on the building blocks $X_\pm$, see \cite{2015arXiv151003836M} for a concrete realization of this idea. In section \ref{sect:2ndconcreteexample} we will use this logic to count the number of bundle moduli on a TCS $G_2$-manifold and confirm a matching with the degrees of freedom of the dual M-theory compactification on a GCS $\Spin(7)$-manifold.

\subsection{Duality for the Building Blocks}
\label{sect:newhetcymg2duals}

Before heading to the full construction of M-theory/$\Spin(7)$ and Heterotic/$G_2$ we first explore the dual pairs that are relevant for each building block. Recall:
\begin{itemize}
\item GCS $\Spin(7)$-manifold $Z$ is built out of an acyl $\CY_4$ building block $Z_-$ and an acyl $G_2$-building block $Z_+$.
\item TCS $G_2$-manifold $J$ is build out of two acyl $\CY_3$, $X_\pm$.
\end{itemize}
We will now construct dual pairs for $\CY_4$ and $G_2$ and correspondingly on the heterotic side $\CY_3$ in sections \ref{sect:z-anddual} and \ref{sect:z+anddual}.
These will then be used as building blocks to construct the GCS/TCS dual pairs. 
As a preparation we first need to construct pairs of $G_2$  and $\CY_3$ and duals for M/het duality.

\subsubsection{Geometric Preparation: Dual Pairs of $G_2$ and $\CY_3$}
\label{sect:newdualg2andcy}

In this section we discuss a well-suited example for building the $G_2$ building block of the GCS-manifold, by consider M-theory on $G_2$-manifold and the heterotic on $\CY_3$ compactification. Similar to the strategy used in \cite{Harvey:1999as}, we can find such a pair by forming an appropriate quotient of a dual pair of 4d ${\mathcal N}=2$ compactifications, which are 
\be\label{MhetN2}
\hbox{M/$X_{43,43}\times S^1$} \quad \longleftrightarrow\qquad  \hbox{Het/K3$\times T_h^2$} \,.
\ee
On the heterotic side we choose a trivial vector bundle on the K3 surface and break $E_8 \times E_8$ to $U(1)^{16}$ by introducing Wilson lines on an the $T_h^2$ factor. Tadpole cancellation then necessitates the introduction of $24$ NS5-branes wrapped on $T_h^2$. 
Such a compactification of the heterotic string theory on K3$\times T_h^2$ is dual to M-theory compactified on $X_{43,43} \times S^1$, where $X_{43,43}$ is a $\CY_3$-fold, chosen to be a fibration of K3-surfaces with polarizing lattice $N = U \oplus (-E_8) \times (-E_8)$ over $\P^1$. The resulting $\CY_3$ is constructed as a hypersurface in a toric variety $A$ associated to the pair of reflexive polytopes
\begin{equation}\label{eq:polyX3prepexample}
\Delta^\circ = \left(\begin{array}{rrrrr}
-1 & 0 & 2 & 2 & 2 \\
0 & -1 & 3 & 3 & 3 \\
0 & 0 & -12 & 0 & 12 \\
0 & 0 & -1 & 1 & -1
\end{array}\right) 
\hspace{1cm}
\Delta = \left(\begin{array}{rrrrr}
-2 & 1 & 1 & 1 & 1 \\
1 & -1 & 1 & 1 & 1 \\
0 & 0 & -1 & 0 & 1 \\
0 & 0 & -6 & 6 & -6
\end{array}\right)
\end{equation}
and has $h^{1,1}(X_{43,43})=h^{2,1}(X_{43,43})=43$. Besides the 4d $\mathcal{N}=2$ supergravity multiplet (which also contains a graviphoton), both of these compactifications have $43$ $\mathcal{N}=2$ vector multiplets, $43$ hypermultiplets and one one universal hypermultiplet. 

We can now quotient both theories in (\ref{MhetN2}) by an involution to find dual compactifications of M-theory on a $G_2$-manifold and heterotic strings on a Calabi-Yau three-fold with a vector bundle. This truncates the light excitations to be described by a theory with $20$ $\mathcal{N}=1$ $U(1)$ vectors and $67$ $\mathcal{N}=1$ chiral multiplets.

On the type IIA side we act with a free involution $\Z_2^{o}$ to produce a $G_2$-manifold\footnote{As the involution we consider acts freely, this produces only a `barely' $G_2$-manifold, the holonomy group of which is only the subgroup $SU(3) \rtimes Z_2$ of $G_2$.} $M$ as
\begin{equation}\label{MDef}
 M = (X_{43,43} \times \mathbb{S}^1_u )/ \Z_2^{o}\, ,
\end{equation}
where $\Z_2^{o}$ sends the coordinate $u\rightarrow -u$ and acts on $X_{43,43}$ as a antiholomorphic involution. This anti-holomorphic involution is chosen such that it sends all of the homogeneous coordinates of the ambient toric space $A$ for $X_{43,43}$ to their complex conjugate and furthermore swaps any homogeneous coordinate of $A$ associated with the lattice point $(a,b,c,d)$ on $\Delta^\circ$ with the homogeneous coordinate associated with the lattice point $(a,b,-c,d)$ on $\Delta^\circ$. Some thought reveals that this is a symmetry of $A$ and that one can choose the hypersurface equation of $X_{43,43}$ such that $\Z_2^{o}$ becomes a symmetry of $X_{43,43}$ (without enforcing singularities) and acts as $\Z_2^{o}: \Omega \rightarrow \bar{\Omega}$, as appropriate for an antiholomorphic involution. There are $47$ divisors on $X_{43,43}$, which descend from $A$ by intersecting $X_{43,43}$ with the vanishing loci of the homogeneous coordinates of $A$, out of which $43$ are linearly independent. There are $21$ pairs of homogeneous coordinates $\{ z_{i\pm}; i=1\cdots 21\}$ on which $\Z_2^{o}$ acts as
\begin{equation}\label{eq:orientifoldaction_part1}
\Z_2^{o} :\qquad z_{i\pm} \rightarrow \pm \bar{z}_{i\mp}
\end{equation}
and the remaining $5$ homogeneous coordinates $\{w_j;j=0\cdots 5\}$ are acted on as
\begin{equation}\label{eq:orientifoldaction_part2}
\Z_2^{o} : \qquad w_j \rightarrow \bar{w}_j \, . 
\end{equation}
Such an involution acts freely, because a fixed point would require that $z_{i\pm} = \pm \bar{z}_{i\mp}$, up to the $\C^*$ action on the homogeneous coordinates of $A$. For our choice of $z_{i\pm}$ the $\C^*$ actions imply that fixed points require $z_{i+}=z_{i-}=0$. For our choice, such intersection vanish for any pair $z_{i\pm}$, so that there are no fixed points. Out of the four linear relations among the divisors of $A$, only one involves the $42$ divisors associated with $\{ z_{i\pm}; i=1\cdots 21\}$ and the remaining three involve the $\{w_j;j=0\cdots 5\}$. Hence there are precisely $20$ even combinations of such divisors and we find that\footnote{Another way to see this is as follows: the three-fold $X_{43,43}$ is fibered by elliptic K3 surfaces with two $II^*$ over $\P^1$ and this involution acts as the antipodal map of the base of this elliptic fibration (another way to see why it acts freely) in particular identifying the two $II^*$ fibers. It furthermore acts (non-freely) on the $\P^1$ base of the K3 fibration and identifies the $24$ reducible K3 fibers on $X_3$ pairwise. This reproduces why there are $8+12=20$ even classes in $b^2(X_3)$.}
\begin{equation}
h^{1,1}_+(X_{43,43}) = 20 \hspace{1cm}  h^{1,1}_-(X_{43,43}) = 23 \, .
\end{equation}
As $\Z_2^{o}$ is an antiholomorphic involution, we also have that $b^3_\pm(X_{43,43})=44$, from which we find
\begin{equation}
\begin{aligned}
 b_2(M) &= h^{1,1}_+(X_3) & = 20 &\\
 b_3(M) &= h^{1,1}_-(X_3) + b^3_+(X_3) & = 67 &  \,,
\end{aligned}
\end{equation}
which gives the number of $\mathcal{N}=1$ $U(1)$ vector and chiral multiplets for compactification of M-Theory on $M$. 

On the dual heterotic side, the corresponding involution $\Z_2^{h}$ acts as a combination of $z\rightarrow -z$ (on the $T_h^2$ factor) together with the Enriques involution\footnote{The Enriques involution is the unique fixed-point free non-symplectic involution on a K3 surface and is identified by the invariants $(r,a,\delta)=(10,10,0)$ in the classification of \cite{Nikulin86discretereflection}.} on the K3 surface to produce a Calabi-Yau three-fold
\begin{equation}\label{HetXDef}
X_{11,11} = (\hbox{K3}\times T_h^2)/\Z_2^h 
\end{equation}
with Hodge numbers $h^{1,1}(X_{11,11}) = h^{2,1}(X_{11,11}) =11$. Furthermore, it acts on the bundle data by twisting the two Wilson lines such as to break $E_8\times E_8$ to $U(1)^8$. We can think about the surviving bundle data as an extension of a line bundle associated with a 
Wilson line breaking $E_8\times E_8$ to $U(1)^8$, and there are eight complex degrees of freedom specifying this bundle.

The orbifold furthermore identifies the $24$ NS5-branes on $T_h^2$ pairwise\footnote{Equivalently, one can compute that $ch_2(X_{11,11})=12 [T_h^2]$.}, so that the number of 4d $\mathcal{N}=1$ $U(1)$ vector and chiral multiplets on the heterotic side is given by
\begin{equation}
\begin{aligned}
n_v &= 8 + 12 &= 20 \\
n_c & = 1 + 2\cdot 11 + 3\cdot 12 + 8 &= 67
\end{aligned}
\end{equation}
in perfect agreement with the dual M-theory on the $G_2$-manifold $M$.
With this preparation in hand we can now study the duality for the building blocks.

\subsubsection{The acyl $G_2$-Manifold and its dual acyl $\CY_3$}
\label{sect:z-anddual}

We can turn $X_{11,11}$ and $M$ in the last subsection 
into a dual pair of an acyl Calabi-Yau three-fold and an acyl $G_2$-manifold by cutting each of them in half. Let us start with $M$ in  (\ref{MDef}). By cutting along the middle of the interval remaining of the circle $\mathbb{S}_u^1$, $M$ is turned into an acyl (barely) $G_2$-manifold $Z_-$ which we may represent by 
\begin{equation}
Z_- = \left( X_{43,43} \times \R_u \right) /  \Z_2^{o} \, ,
\end{equation}
where $\Z_2^{o} $ acts as before on $X_{43,43}$ and as $u\rightarrow -u$ on $\R_u$. 
The asymptotic Calabi-Yau three-fold of $Z_-$ is $X_{43,43}$ and there are $b^2_+(X_{43,43})=20$ classes in the image of the restriction map from $H^2(Z_-,\Q)$ to $H^2(X_{43,43},\Q)$, as well as $b^3_+(X_{43,43})=44$ classes in the image of the restriction map from $H^3(Z_-,\Q)$ to $H^3(X_{43,43},\Q)$. There are no elements in the kernels of these maps. 

On the heterotic side, Calabi-Yau three-fold $X_{11,11}$ (\ref{HetXDef}) carries a natural (almost trivial) K3 fibration 
\be
\hbox{K3} \hookrightarrow  X_{11,11} \rightarrow \mathbb{P}^1_h\,,
\ee
where the base $\mathbb{P}^1_h$ is found as the quotient of $T^2_h$. Over four points on this $\P_h^1$, the K3 fiber is truncated to an Enriques surface, but is constant otherwise. Note that this way of thinking reproduces the fact that the Euler characteristic of $X_{11,11}$ vanishes from
\begin{equation}
\chi(X_{11,11}) = 24\cdot(2-4) + 4\cdot 12 =0 \, ,
\end{equation}
where we have excised the four fibers which are Enriques surfaces from the base $\P_h^1$ in the first term and added them back in for the second. If we now cut $\P_h^1$ into two halves, each of which is $\C$ and contains two of the Enriques fibers over it, we produce an acyl Calabi-Yau three-fold $X_-$ as the TCS $G_2$ building block 
\begin{equation}
\begin{aligned}
K(X_-) &=  \emptyset \\
N(X_-) &= U(2) \oplus (-E_8(2)) \,,
\end{aligned}
\end{equation}
where $L(2)$ denotes a lattice $L$ with its inner form rescaled by a factor of two. The expression for $N(X_-)$ is nothing but the even sublattice of $H^2(K3,\mathbb{Z})$ under the Enriques involution. Furthermore, $X_-$ may be compactified to a compact building block $\tilde{X}_-$ in the sense of \cite{MR3109862} by gluing in a single K3 fiber. It now follows from a computation as above that $\chi(\tilde{X}_-)=24$, which together with $b^2(\tilde{X}_-) = 11$ (and $b^1(\tilde{X}_-)=0$ implies that $b^3(\tilde{X}_-)=0$. This is all of the data we will need in using $X_-$ as a building block for a $G_2$-manifold. 

Note that we have indeed decomposed $M$ and $X_{11,11}$ by cutting along the same $\mathbb{S}^1$. 
This becomes particularly clear, by noting that these dual compactifications descend from a compactification to five dimension, in which M-theory is put on $X_{43,43}$ and heterotic string on K3$\times S^1$. The dual models we have constructed are then found by compactifying both side on a further $\mathbb{S}_u^1$ and quotienting by $\Z_2$. It is the quotient of $\mathbb{S}_u^1$, which becomes one of the two circles on $T^2_h$, which is cut in half to produce $Z_-$ and $X_-$, respectively. 

\subsubsection{The acyl $\CY_4$ and its dual acyl $\CY_3$}\label{sect:z+anddual}

We now construct a dual pair of an acyl Calabi-Yau four-fold $Z_+$ (as a building block for the $\Spin(7)$ GCS) and the heterotic dual  acyl Calabi-Yau three-fold $X_+$ (as a building block for the $G_2$ TCS). As we want $Z_+$ to asymptote to $X_{43,43}$ so that we can glue $Z_+$ with $Z_-\times S^1$, we construct 
\be
Z_+ = \tilde{Z}_+ \setminus X_{43,43}^0
\,, \qquad X_{43,43} \hookrightarrow \tilde{Z}_{+} \rightarrow \mathbb{P}^1\,,
\ee
so that the algebraic three-fold $\tilde{Z}_+$ satisfies
\begin{equation}
c_1( \tilde{Z}_+) = [X_{43,43}^0] \, .
\end{equation}
Note that we may also construct $Z_+$ by cutting a Calabi-Yau four-fold $X_4$, which is a fibration of $X_{43,43}$ over $\P^1$ in two halves. This four-fold is a toric hypersurface defined by the pair 
\begin{equation}\label{eq:firstexdoubecoverCY4}\ba
\Delta^\circ &= \left(\begin{array}{rrrrrrrr}
-1 & 0 & 2 & 2 & 2 & 2 & 2 & 2 \\
0 & -1 & 3 & 3 & 3 & 3 & 3 & 3 \\
0 & 0 & -12 & -12 & 12 & 12 & 0 & 0 \\
0 & 0 & -1 & 0 & 0 & -1 & 0 & 1 \\
0 & 0 & 0 & 1 & 1 & 0 & -1 & 0
\end{array}\right) \cr 
\Delta &= \left(\begin{array}{rrrrrrr}
-2 & 1 & 1 & 1 & 1 & 1 & 1 \\
1 & -1 & 1 & 1 & 1 & 1 & 1 \\
0 & 0 & -1 & 0 & 0 & 0 & 1 \\
0 & 0 & -6 & -6 & 6 & 6 & -6 \\
0 & 0 & 6 & -6 & -6 & 6 & 6
\end{array}\right)
\ea\end{equation}
of reflexive polytopes. Its Hodge numbers are
\begin{equation}
h^{1,1}(X_4) = 68\, ,\hspace{.5cm} h^{3,1}(X_4) = 292 \, ,\hspace{.5cm} h^{2,1}(X_4) = 0 \, ,\hspace{.5cm} h^{2,2}(X_4) =1484 \, ,
\end{equation}
so that $\chi(X_4) = 2208$. As a generalization of \cite{Braun:2016igl} to four-folds, $\tilde{Z}_+$ can also be constructed from one of the two isomorphic tops in $\Delta^\circ$ corresponding to the fibration by $X_{43,43}$. Either way, it follows that the Euler characteristic of $Z_+$ is $\chi(Z_+) = 1104$, half of the Euler characteristic of $X_4$.

The restriction maps $\beta^i:H^i(Z_+,\Q) \rightarrow H^i(X_{43,43}^0\times \bbS^1,\Q)$ are fairly easy because all of the divisors and none of the three-cycle on $X_{43,43}$ descend from $Z_+$. Furthermore, there are $12$ divisor classes on $Z_+$ associated with reducible fibers of the fibration by $X_{43,43}$. Hence we conclude
\begin{equation}
\begin{aligned}
|\im(\beta^2)| &= 43 \, ,  \hspace{.5cm} &|\ker(\beta^2)| &= 12 \, ,  \hspace{.5cm} &|\coker(\beta^2)| &= 0 \\ 
|\im(\beta^3)| &= 0 \, ,  \hspace{.5cm} &|\ker(\beta^3)| &= 0 \, ,  \hspace{.5cm} &|\coker(\beta^3)| &= 131 \\  
\end{aligned}
\end{equation}
This will be all the data from $Z_+$ we need in the following. 


Let us now find the dual geometry for the heterotic string. As $Z_+$ and $X_4$ are fibered by elliptic K3 surfaces, this is fairly straightforward, as we simply need to apply the known rules of heterotic-F-theory duality. In particular, $Z_+$ and $X_4$ are fibered by elliptic K3 surfaces with Picard lattice
$U\oplus (-E_8)^2$, so that the dual heterotic theory has no $E_8 \times E_8$ vector bundles turned on. Furthermore, $Z_+$  and $X_4$ are elliptic fibrations over (blow-ups of) $\C \times \P^1 \times \P^1$. These blowups capture the presence of the NS5-branes needed on the heterotic side to satisfy the Bianchi identity. Hence the dual heterotic geometry dual to is $X_4$ given by an elliptically fibered Calabi-Yau $X_{3,243}$ with base $\P^1 \times \P^1$ times a circle $\bbS^1$. The three-fold $X_{3,243}$ is constructed from a pair of reflexive polytopes with vertices
\begin{equation}\label{eq:polytopesx3,243}
 \Delta^\circ = \left(\begin{array}{rrrrrr}
-1 & 0 & 2 & 2 & 2 & 2 \\
0 & -1 & 3 & 3 & 3 & 3\\
0 & 0 & -1 & 1 & 0 & 0\\
0 & 0 & 0 & 0 & 1 & -1
\end{array}\right) \, ,\hspace{.5cm}
\Delta = \left(\begin{array}{rrrrrrrr}
-2 & 1 & 1 & 1 & 1 & 1 &1&1\\
1 & 1 & 1 & 1 & 1 & -1 &1&1\\
0 & 6 & 6 & -6 & -6 & 0 &6&-6\\
0 & 0 & -6 & 0 & -6 & 0 &6&6
\end{array}\right)\, .
\end{equation}
and it follows that $X_{3,243}$ has Hodge numbers $(h^{1,1},h^{2,1})=(3,243)$. The dual acyl CY building block $X_+$ of the acyl Calabi-Yau four-fold $Z_+$ is constructed using by cutting $X_{3,243}$ along an $S^1$ in the base of its K3 fibration. Equivalently, we may glue in a K3 fiber to compactify $X_+$ to a building block $\tilde{X}_+$ which can be constructed using the methods of \cite{Braun:2016igl} from a pair of projecting tops
\begin{equation}
 \Diamond^\circ = \left(\begin{array}{rrrrr}
-1 & 0 & 2 & 2 & 2 \\
0 & -1 & 3 & 3 & 3 \\
0 & 0 & -1 & 1 & 0 \\
0 & 0 & 0 & 0 & 1
\end{array}\right) \, ,\hspace{.5cm}
\Diamond = \left(\begin{array}{rrrrrr}
-2 & 1 & 1 & 1 & 1 & 1 \\
1 & 1 & 1 & 1 & 1 & -1 \\
0 & 6 & 6 & -6 & -6 & 0 \\
0 & 0 & -6 & 0 & -6 & 0
\end{array}\right)\, .
\end{equation}
It follows that 
\begin{equation}
\begin{aligned}
N(X_+) &= U \\
K(X_+) &= \emptyset \\
h^{2,1}(\tilde{X}_+) &= 112 \, .
\end{aligned}
\end{equation}
Note that there are Wilson lines breaking $E_8\times E_8$ to $U(1)^{16}$ on the product $\bbS^1$ on the heterotic side, which are dual to the volumes of the divisors associated with the $(-E_8)^{\oplus 2}$ in the Picard lattice of the K3 fibers of $X_4$.

\subsection{First Example of Dual Pairs: New $\Spin(7)$- and its dual $G_2$-Manifold}\label{sect:example1}

Above we have constructed a pair of an acyl Calabi-Yau four-fold $Z_+$ and an acyl $G_2$-manifold $Z_-$ which both asymptote to (a cylinder/an interval times) the same Calabi-Yau three-fold $X_{43,43}$, together with the two dual acyl Calabi-Yau three-folds $X_+$ and $X_-$ together with the corresponding bundle data on the heterotic side. We will now glue $Z_+$ with $Z_-\times \mathbb{S}^1$ to a GCS $\Spin(7)$-manifold $Z$ and $X_\pm\times \mathbb{S}_1$ to a TCS $G_2$-manifold $J$ and verify that the light fields in the effective field theories indeed agree.

\subsubsection{The M-theory on GCS $\Spin(7)$-Manifold}

As detailed in Section \ref{sect:generalconstruction}, the $\Spin(7)$-manifold $Z$ is constructed as
\be
Z = Z_+ \cup (Z_-\times S^1) \,,
\ee
where $Z_+ \cap ( Z_- \times S^1) = X_3 \times S^1=$ and $X_3=X_{43,43}$. It immediately follows that $\chi(Z) =\chi(Z_+) = 1104$. The remaining Betti numbers are found from the Mayer-Vietoris sequence. First,
\be
b^2(Z) = |\ker \gamma^2| + |\hbox{coker} \gamma^1| = (20  + 12 )  + 0 = 32 \,,
\ee
where we have used that $20$ classes in $h^{1,1}(X_{43,43})$ are in the common image of the restriction maps $\gamma^2:H^2(Z_+)\oplus H^2(Z_-\times \mathbb{S}^1)\rightarrow H^2(X_{43,43}\times S^1)$ and that the kernel of $\beta^2_+:H^2(Z_+)\rightarrow H^2(X_{43,43})$ is 12-dimensional. Furthermore, 
\be
b^3(Z) = |\ker \gamma^3| + |\hbox{coker} \gamma^2| = 0 \, ,
\ee
as all of the classes in $h^{1,1}(X_{43,43})$ are in the image of $\beta^2_+$, no class in $b^3(X_{43,43})$ is in the image of $\beta^3_+$ and the restriction maps $\beta^3_\pm$ have no kernels. 

We can now use \eqref{eq:specMspin7} to find
\begin{equation}
 b^4_-(Z) = 327
\end{equation}
and it follows that there are $n_c= 360$ {real degrees of freedom originating from the metric and three-form for compactifications of M-theory on $Z$. Out of these, $32$ sit in 3d $\mathcal{N} = 1$ vector multiplets.} Furthermore, a consistent compactification requires the introduction of $1104/24 = 46$ M2 branes. Each of these contributes a further $8$ real moduli from moving on $Z$. 

{Alternatively, this $\Spin(7)$-manifold can also be obtained as a free quotient of the Calabi-Yau four-fold $X_4$ specified as a toric hypersurface by \eqref{eq:firstexdoubecoverCY4}.}

\subsubsection{The Dual Heterotic Model on TCS $G_2$-Manifold}\label{eq:heteroticmodel1}

The dual heterotic side has various components: geometry, vector bundle and NS5-branes. We will determine all the dual data, and compare with the spectrum obtained in the M-theory compactification -- and find agreement. 

\subsubsection*{Geometry}

M-theory on $Z$ is dual to heterotic on a $G_2$-manifold $J$ which can be obtained by fiber-wise duality as 
\be
J= (X_+ \times \bbS^1_{e,+} ) \cup (X_- \times  \bbS^1_{e,-} ) 
\ee
where $X_\pm$ are the acyl Calabi-Yau three-folds introduce in sections \ref{sect:z+anddual} and \ref{sect:z-anddual} above. Recall that the lattices $N(X_\pm)$ are
\begin{equation}
\begin{aligned}
N(X_+) &= U\\
N(X_-)& = U(2) \oplus E_8(2) \,,
\end{aligned}
\end{equation}
and there is a matching such that $N(X_+)\cap N(X_-) = \emptyset$. Together with $K(X_\pm)$ and $h^{2,1}(X_\pm)$ we hence immediately find \cite{Corti:2012kd}
\begin{equation}
\begin{aligned}
b_2(J) &= K(X_+) + K(X_-) = 0\\
b_2(J)  + b_3(J) &= 23 + 2\left(K(X_+) + K(X_-)\right) + 2\left(h^{2,1}(X_+)+h^{2,1}(X_-)\right) = 247 \, .  
\end{aligned}
\end{equation}
giving $b_3(J)=247$. We hence find $247$ real degrees of freedom from the metric of a heterotic compactification on $J$ and no moduli associated with the $B$-field.

\subsubsection*{Bundle data}

Recall from section \ref{sect:newdualg2andcy} that the heterotic compactification on $X_{11,11} = X_- \cup X_-$ dual to M-theory on $M = Z_- \cup Z_-$ has eight complex bundle degrees of freedom, associated with a bundle which is an extension of appropriate Wilson lines on the double cover of $X_{11,11}$. The origin of the corresponding degrees of freedom in the dual M-theory suggests to associate $4$ complex bundle degrees of freedom with each of the $S^1$s in the $K3 \times T^2_h$ double cover of $X_{11,11}$. For $X_-$, one of these two $S^1$s has disappeared, so that we are left with 8 real bundle degrees of freedom there. In particular, these give the non-trivial holonomies in $E_8\times E_8$ along $\bbS^1_{b,-}$ in the base $\C$ of the K3 fibration on $X_-$. In the TCS $G_2$-manifold, these get mapped to Wilson lines on $\bbS^1_{e,}$. Note that this is precisely what is expected for the duality between M-theory on the Calabi-Yau four-fold $X_4$ and heterotic strings on $X_{3,243}\times \bbS^1_e$: the non-zero volumes of the divisors on $Z_+$ associated with the $(-E_8)^{\oplus 2}$ in the Picard lattice of the K3 fibers of $Z_+$ are mapped to Wilson lines on $\bbS^1_{e+}$ on the heterotic side. We hence conclude that $J$ carries a vector bundle $V$ inherited from a bundle on $X_-$ which breaks $E_8\times E_8\rightarrow U(1)^8$ and has eight real moduli. The data of this bundle is captured by the volumes of the curves in $(-E_8^{\oplus 2})$ of the K3 fibers of $Z_\pm$.

\subsubsection*{NS5-branes}

Furthermore, our compactifications of the heterotic string require the addition of NS5-branes for consistency as the gauge bundle $V$
does not have a second Chern character. We will infer the NS5-branes which we need to include by exploiting piecewise duality. In particular, we can think about a representative of the first Pontrjagin class of $J$ as being glued from representatives of $c_2 (\tilde{X}_\pm)$. 

The second Chern class of $\tilde{X}_+$ is 
\begin{equation}
 \begin{aligned}
  c_2 (\tilde{X}_+) & = 46 H \hat{H}_+ + 46 H_+ \sigma_+ + 23 \hat{H}_+ \sigma_+ + 11 \sigma_+^2 \\
  &= 46 H_+\hat{H}_+ + 24 H_+ \sigma_+ + 12 \hat{H}_+ \sigma_+\,,
 \end{aligned}
\end{equation}
where $\hat{H}_+$ is the divisor class corresponding to fixing a point on the base of the K3 fibration on $\tilde{X}_+$ and $H_+$ is the class corresponding to fixing a point on the other $\P^1$ in the $\P^1 \times \hat{P}^1$ base of the elliptic fibration on $\tilde{X}_+$. The divisor class $\sigma_+$ is represented by the section of this elliptic fibration and these classes satisfy the relation $\sigma_+ (\sigma_+ + 2 H_+ + \hat{H}_+) =0$. Note that $c_1(\tilde{X}_+) = [\hat{H}_+]$. 

The $46$ NS5-branes on the curve $H\hat{H}$ are wrapped on the elliptic fiber of $X_+$ times the auxiliary $\bbS^1_{+,e}$ multiplying $X_+$ in the construction of $J$. As such, they are wrapped precisely on the conjectured fiber of the $T^3$ fibration of $J$ used for the duality and we can associated them with the $46$ M2 branes present on the M-theory side. Let us confirm this by counting the number of moduli. Each of these NS5-branes has $4$ real degrees of freedom by displacing the elliptic fiber on $X_+$. Furthermore, there is one real degree of freedom from the scalar $\phi$ in the tensor multiplet (the position along the 11th direction in heterotic M-theory) and three real degrees of freedom from the self-dual two-form $B$ on the $b^1(T^3)=b^2(T^3)=3$ cycles of the $T^3$ they are wrapped on. This makes 8 degrees of freedom matching the counting for the M2 branes. 

The $12$ NS5-branes on $\hat{H}_+ \sigma_+$ are points on the $\P^1$ base of the K3 fibration of $\tilde{X}_+$ so that they are wrapped on a three-manifold with the topology $S^2 \times S^1$ in $J$. Within $X_+$, each such NS5-brane has $2$ real moduli associated with displacement\footnote{The associated holomorphic curves which lift to associatives are fixed to lie on the section of the elliptic fibration of $X_+$}, one modulus associated with the worldvolume scalar $\phi$, and one modulus from the self-dual two-form $B$. At least in the Kovalev limit, we are hence led to associate $4$ real moduli with each such NS5-brane. 

Finally, there are the $24$ NS5-branes in the class $H_+ \sigma_+$ of $\tilde{X}_+\times \bbS^1_{e,+}$. On $X_+ = \tilde{X}_+\setminus S_+^0$, they are wrapped on the whole of the open base $\C \times \bbS^1_{e,+}$ of the K3 fibration on $X_+ \times \bbS^1_{e,+}$. Similarly, there are $12$ NS5-branes wrapped on a double cover the open base $\C \times \bbS^1_{e,-}$ of the K3 fibration on $X_- \times \bbS^1_{e,-}$ as we have seen in Section \ref{sect:newdualg2andcy}. Note that this sector of NS5-branes becomes $24$ copies of the $T^2$ factor times the interval in $(X_+ \times \bbS^1_{e,+} ) \cap (X_- \times  \bbS^1_{e,-} ) =  I \times T^2 \times K3$. On the $G_2$-manifold, these NS5-branes can hence be joined to form $12$ irreducible NS5-branes.  This has several effects. First of all, the relative positions of the $24$ branes on $X_+\times\bbS^1_{+,e}$, which are points on the K3 fiber $S_+$ of $X_+\times$, are pairwise fixed to be symmetric under the Enriques involution acting on $S_-$. Second, each such pair of NS5-branes only has two real moduli of deformation. Because they are wrapped on the $\C$-base of the K3 fibration on $X_+$ and the elliptic fibration of $X_+$ is non-trivial there, they cannot be displaced in the direction of the elliptic curve of $X_+$. Third, the two branches of each pair of NS5-branes are swapped when encircling two special points in the base of the K3 fibration on $X_-$, so that we should think of each such pair of NS5-branes as a single NS5-brane wrapped on a three-manifold $L$ which is a double cover of $S^3$ branched along two unlinked $S^1$s. We conclude that there are $12$ NS5-branes wrapped on three-manifolds $L$ inside $J$ with two real deformations each. It is not hard to see that $b^1(L) = b^2(L) = 1$, so that each of these NS5-branes is associated with 4 real degrees of freedom: $2$ real moduli of displacement together with $2$ real moduli from $\phi$ and $B$ adding up to $4$ real moduli each.


\subsubsection*{Counting of Degrees of Freedom}

We are now ready to count the number of degrees of freedom on the heterotic side. To do so, we can neglect the $46$ NS5-branes wrapped on the $T^3$ fiber of $J$, as these are mapped to M2 branes in M-theory and we have already matched their degrees of freedom. Summarizing the different contributions discussed above, the remaining light fields for heterotic strings on $J$ are hence the $247$ moduli of the Ricci-flat metric on $J$, the $8$ real degrees of freedom in the bundle $V$, $8$ $U(1)$ vectors which remain massless in $E_8\times E_8$, $24$ NS5-branes with 4 real moduli each, and the dilaton, giving a total of $360$ real degrees of freedom. Out of these, there are $32$ 3d $\mathcal{N} = 1$ vector multiplets, originating from the $8$ surviving $U(1)$s together with the $24$ NS5-branes wrapped on three-cycles with $b^1=1$. We have hence verified that the number of light fields between heterotic string theory on $J$ and M-theory on $Z$ precisely agrees. 


\subsection{Second Example of Dual Pairs}\label{sect:2ndconcreteexample}

Let us now make consider a variation of the previous example and work out the topology of the $\Spin(7)$ associated with putting an $E_8\times E_8$ vector bundle on $J$. To describe such a situation, we intend to replace $Z_-$ and $Z_+$ by an acyl $G_2$-manifolds and and acyl $\CY_4$ originating from a K3 fibration with Picard lattice $U$ instead of $U\oplus (-E_8)^{\oplus 2}$.

\subsubsection{The acyl Calabi-Yau four-fold $Z_+$}\label{sect:acyl4foldexample2}

Following the usual rules of heterotic-M-theory (F-theory) duality $Z_+$ is now found as one half of a compact Calabi-Yau four-fold $X_4$ described by a generic Weierstrass elliptic fibration over $\P^1\times\P^1\times\P^1$ and $Z_+ \cap (Z_- \times \bbS^1) = X_{3,243} \times \bbS^1 \times I$. Here, $X_{3,243}$ has already appeared in Section \ref{sect:z+anddual} and $X_4$ is found from the pair of reflexive polytope:
\begin{equation}\label{eq:ex2doublecovercy4}\ba
 \Delta^\circ &= \left(\begin{array}{rrrrrrrr}
-1 & 0 & 2 & 2 & 2 & 2 & 2 & 2 \\
0 & -1 & 3 & 3 & 3 & 3 & 3 & 3 \\
0 & 0 & -1 & 0 & 0 & 1 & 0 & 0 \\
0 & 0 & 0 & -1 & 0 & 0 & 0 & 1 \\
0 & 0 & 0 & 0 & -1 & 0 & 1 & 0
\end{array}\right)\cr 
\Delta &=  \left(\begin{array}{rrrrrrrrrr}
-2 & 1 & 1 & 1 & 1 & 1 & 1 & 1 & 1 &
1 \\
1 & -1 & 1 & 1 & 1 & 1 & 1 & 1 & 1 &
1 \\
0 & 0 & -6 & -6 & -6 & -6 & 6 & 6 & 6
& 6 \\
0 & 0 & -6 & -6 & 6 & 6 & -6 & -6 & 6
& 6 \\
0 & 0 & -6 & 6 & -6 & 6 & -6 & 6 & -6
& 6
\end{array}\right)\, ,
\ea\end{equation}
which implies 
\begin{equation}
h^{1,1}(X_4) = 4\, ,\hspace{.5cm} h^{3,1}(X_4) = 2916 \, ,\hspace{.5cm} h^{2,1}(X_4) = 0 \, ,\hspace{.5cm} h^{2,2}(X_4) = 11724 \, ,
\end{equation}
and $\chi(X_4) = 17568$. We can write $Z_+ \cup Z_+ = X_4$ with $Z_+ \cap Z_+ = X_{3,243} \times \bbS^1 \times I$ and the data of $Z_+$ relevant for its use in the construction of $Z$ are
\begin{equation}
\begin{aligned}
|\im \beta^2_+| &= 3 \, ,\hspace{1cm}  &|\ker \beta^2_+| &= 0 \\
|\im \beta^3_+| &= 0 \, ,\hspace{1cm}  &|\ker \beta^3_+| &= 0 \\
\end{aligned}
\end{equation}
and $\chi(Z_+) = 8784$.

\subsubsection{The acyl $G_2$-manifold $Z_-$}

To construct $Z_-$, we first consider $M = Z_- \cup Z_- = X_{3,243} \times \bbS^1_u / \Z_2$. The three-fold $X_{3,243}$ was constructed as a toric hypersurface associated with the pair of reflexive polytopes shown in \eqref{eq:polytopesx3,243}. Its homogeneous coordinates have the toric weights
\begin{equation}
 \begin{array}{ccccccc}
  y & x & w & z_1 & z_2 & \hat{z}_1 & \hat{z}_2 \\
  \hline
  3 & 2 & 1 & 0 & 0 & 0 & 0\\
  6 & 4 & 0 & 1 & 1 & 0 & 0 \\
  6 & 4 & 0 & 0 & 0 & 1 & 1
 \end{array}\, .
\end{equation}
Similar to our procedure in section \ref{sect:newdualg2andcy} we can introduce an antiholomorphic involution on $X_{3,243}$ by letting\footnote{In fact, $X_{43,43}$ is connected to $X_{3,243}$ by several singular transitions and the actions \eqref{eq:orientifoldaction_part1} and \eqref{eq:orientifoldaction_part2} imply the one given here.}
\begin{equation}
\begin{aligned}
(y,x,w,\hat{z}_1,\hat{z}_2) &\rightarrow  (\bar{y},\bar{x},\bar{w},\hat{\bar{z}}_1,\hat{\bar{z}}_2) \\
(z_1,z_2) &\rightarrow  (-\bar{z_2},\bar{z}_1) \, .
\end{aligned}
\end{equation}
This involution acts as the antipodal map on the $\P^1$ with homogeneous coordinates $[z_1,z_2]$ and hence acts freely on $X_{3,243}$ such that
\begin{equation}
\begin{aligned}
h^{1,1}_+(X_{3,243})& = &0 &\hspace{1cm}& h^{1,1}_-(X_{3,243})& =& 3 \\
h^{2,1}_+(X_{3,243}) &= & 244 && h^{2,1}_-(X_{3,243})& =& 244 
\end{aligned}
\end{equation}
The resulting barely $G_2$-manifold $M$ hence has Betti numbers 
\begin{equation}
b^2(M) = 0\, ,\hspace{1cm} b^3(M) = 247 \, . 
\end{equation}
We can now form an acyl $G_2$-manifold $Z_-$ by cutting $\bbS^1_u$ in the middle, i.e. $M = Z_- \cup Z_-$ and $Z_- \cup Z_- = \bbS^1 \times X_{3,243}$. The restriction maps are
\begin{equation}
\begin{aligned}
|\im \beta^2_-| &= 0 \, ,\hspace{1cm}  &|\ker \beta^2_-| &= 0 \\
|\im \beta^3_-| &= 244 \, ,\hspace{1cm}  &|\ker \beta^3_-| &= 0 \,.
\end{aligned}
\end{equation}

\subsubsection{M-theory Spectrum on the GCS $\Spin(7)$-Manifold}

From the acyl Calabi-Yau four-fold $Z_+$ and the acyl $G_2$-manifold $Z_-\times \bbS$ we can now form a $\Spin(7)$-manifold $Z=Z_+ \cup Z_-\times \bbS$. It follows from the restriction maps that 
\begin{equation}
b^2(Z)=b^3(Z) = 0\, .
\end{equation}
Furthermore $\chi(Z) =  8784$. Hence 
\begin{equation}
b^4_-(Z) = 2919 
\end{equation}
and compactifications of M-theory on $Z$ have $2920$ real degrees of freedom from the metric which all sit in real multiplets, as well $8784/24=366$ space-time filling M2-branes. 

Again, this $\Spin(7)$-manifold can also be obtained as a free quotient of the Calabi-Yau four-fold $X_4$ specified as a toric hypersurface by \eqref{eq:ex2doublecovercy4}.

\subsubsection{The Dual Heterotic Model}

The dual heterotic compactification lives on the same $G_2$-manifold $J$ as discussed in Section \ref{eq:heteroticmodel1}. Here, the metric contributes $247$ real degrees of freedom. Now, however, we choose to switch on a vector bundle $V$ completely breaking $E_8 \times E_8$, such that $V$ is flat on the $T^3$ fibers of $J$ for the duality to hold fiberwise. It now follows from the proposed duality that such a bundle must have
\begin{equation}
m_V = 2920 - 247 -1 =  2672
\end{equation}
real moduli. Furthermore, the heterotic Bianchi identity in this context must force to include $366$ NS5-branes wrapped on the $T^3$ fiber of $J$.

We can reproduce the number of bundle moduli by studying the moduli of holomorphic vector bundles on $X_\pm$ together with a gluing condition. For $X_-$, such bundles are inherited from $E_8 \times E_8$ vector bundles on K3 symmetric under the Enriques involution. From the point of view of the K3 fibration on $X_-$, the bundle data are constant over the base $\C$ of $X_-$. A holomorphic vector bundle on K3 has $4\cdot 112$ real moduli (they all sit in hyper-multiplets for heterotic string theory on K3) and the restriction to be invariant imposes a restriction fixing half, i.e. $224$ real moduli of such a bundle. 

For $X_+$, the number of bundle moduli can be inferred as follows. Consider heterotic strings on $X_{3,243} = X_+ \cup X_+$. By a straightforward application of duality to F-theory (which is compactified on the manifold $X_4$ discussed in Section \ref{sect:acyl4foldexample2}) immediately gives that a generic vector $E_8 \times E_8$ model $W$, which is flat on the elliptic fiber of $X_{3,243}$, has $m_W = 5344$ real moduli \cite{Friedman:1997yq,Andreas:1997ce}. An $E_8 \times E_8$ vector bundle on $X_+ \cap X_+ = I \times \bbS^1 \times K3$ has $4\cdot 112$ real moduli, which gives the number of conditions when matching two generic $E_8 \times E_8$ bundles $W_\pm$ on $X_\pm$. We hence expect
\begin{equation}
m_{W_+} + m_{W_-} - 448 = m_V  \, .
\end{equation}
As furthermore $m_{W_+} = m_{W_-}$, we find that $m_{W_+} = 2896$. 

On the $G_2$-manifold $J$, we can hence construct a suitable vector bundle $V$ by letting $V|_{X_+\times \bbS^1_{e,+}} = W_+$. Such a bundle restricts to a vector bundle on $S_+$ which then has to be appropriately restricted to match $V|_{X_-\times \bbS^1_{e,-}}$. As we have seen $V_{S_-}$ is not a generic vector bundle on a K3 surface, but there are $224$ conditions arising at it must be symmetric under the Enriques involution. Once this condition is met, $V_{S_-}$ uniquely defines $V|_{X_-\times \bbS^1_{e,-}}$. As we have seen, the Enriques involution gives $224$ real restrictions, so that we conclude that
$m_V = 2896 - 224 = 2672$, which matches the expectation from the counting of degrees of freedom on the $\Spin(7)$-manifold $Z$. {Ignoring the 366 NS5-branes wrapped on the $T^3$ fiber of $J$, all of the moduli of this model sit in real multiplets, matching the result that $b^2(Z)=0$ on the M-Theory side. }

\section{Discussion and Outlook}
\label{sec:Outlook}

We proposed a new construction of eight-manifolds with $\Spin(7)$-holonomy, based on a generalized connected sum (GCS), where two building blocks -- a Calabi-Yau four-fold and a $G_2$-holonomy manifold times $S^1$ -- are glued together along an asymptotic region that is a Calabi-Yau three-fold times a cylinder. This construction is in part inspired by the recent twisted connected sum (TCS) realization of $G_2$-holonomy manifolds, which has resulted in a multitude of new examples of compact $G_2$-manifolds, and thereby a resurgence of interest in the string/M-theory context. Likewise the GCS-construction that we propose, provides an avenue to construct large classes of compact $\Spin(7)$-manifolds systematically. In particular the construction of acyl Calabi-Yau three-folds using {semi-Fano three-folds \cite{MR3109862}} or tops \cite{Braun:2016igl} which is useful for TCS-constructions, has an obvious generalization to acyl Calabi-Yau four-folds, which is useful for expanding the set of examples of GCS-constructions. 

We gave an alternative description of the GCS $\Spin(7)$-manifolds in terms of a quotient by an anti-holomorphic involution of a Calabi-Yau three-fold fibered Calabi-Yau four-fold in section \ref{sec:CY34}, which is similar in spirit to the constructions of Joyce, however the key difference is that instead of gluing in ALE-spaces at point-like orbifold singularities, we glue in $G_2$-manifolds with suitable asymptotics. 

There is a multitude of future directions to consider: 

\begin{enumerate}
\item M/F-Duality for $\Spin(7)$: M-theory on $\Spin(7)$ results  3d $\mathcal{N}=1$ theories, and we have seen that there are subsectors of the effective field theory, which in the limit of infinite asymptotic region enjoy enhanced supersymmetry, as discussed in section \ref{sec:FieldTheory}. It would clearly be very interesting to apply M/F-duality to the GCS $\Spin(7)$-manifolds and determine how the supersymmetry breaking in the 4d F-theory vacuum is realized \cite{Vafa:1996xn}. Needless to say it is not difficult to construct GCS-examples that have an elliptic fibration and we will return to this shortly elsewhere. Again key for this will be also to understand the fluxes in the GCS $\Spin(7)$-manifolds.

\item M/Heterotic-Duality and Heterotic $G_2$-systems: 
As was exemplified in section \ref{sec:MHet}, some GCS-constructions have a K3-fibration, so that M-theory compactification on these has a dual description in terms of Heterotic on $G_2$-manifolds with a $T^3$-fibration with vector bundle. We have seen that the GCS-decomposition of the $\Spin(7)$-manifold maps in the dual $G_2$-compactification to a TCS-decomposition. Heterotic on $G_2$-manifolds has been studied only very sparsely, and this approach may very well provide further insight into the construction and moduli of vector bundles for so-called heterotic $G_2$-systems.  Furthermore, this duality also gives evidence for the existence of associative $T^3$-fibrations on $G_2$-manifolds as conjecture from $G_2$ mirror symmetry in \cite{Braun:2017csz}, and these may indeed be less rare as was proposed in \cite{joyce2000compact} despite the obstructions of associatives.  

\item Non-abelian gauge groups: Furthermore we expect, again through the duality to heterotic, that the GCS-$\Spin(7)$ manifolds can give rise to non-abelian gauge symmetries. Locally, these will have a description in the form of an ADE singularities over Cayley four-cycle (for a discussion of such non-compact examples see \cite{Gukov:2001hf}). It would indeed be interesting to develop this further and understand e.g. a Higgs bundle description of the effective theories and its relation to the compact $\Spin(7)$ manifolds -- much like what has been done in recent years for M/F-theory on $\CY_4$. 

\item Mirror Symmetry: Mirror symmetry for TCS-$G_2$ manifolds as studied in \cite{Braun:2017ryx,Braun:2017csz} by applying the mirror map to the Calabi-Yau three-fold building blocks. It would be interesting to see whether there is a similar way to study mirror symmetry for $\Spin(7)$ manifolds (as proposed in \cite{Shatashvili:1994zw}) by applying the mirror map to the building blocks. 

\end{enumerate}


\subsection*{Acknowledgements}

We thank for Sebastjan Cizel, Magdalena Larfors, Xenia de la Ossa, James Sparks and Yuuji Tanaka for discussions related to the present work. APB and SSN are supported by the ERC Consolidator Grant 682608 ``Higgs bundles: Supersymmetric Gauge Theories and Geometry (HIGGSBNDL)''.

\appendix

\section{TCS-Construction of $G_2$-Manifolds}
\label{app:TCS}

The TCS $G_2$-manifolds  \cite{MR2024648, MR3109862, Corti:2012kd} have the topology of a K3-fibration over an $S^3$ and are obtained by gluing two building blocks $Z_\pm$ that are algebraic three-folds, which are K3-fibered over $S^2$. We will denote the generic K3-fibers of each building block by $S_\pm$. {Excising a fiber, the two building blocks asymptote to} K3$\times S^1 \times S^1\times I$, where $I$ is an interval.  
The building blocks are not Calabi-Yau but have $c_1 (Z_\pm) = [\,S_\pm\,]$, which means that removing a fiber gives rise to acyl Calabi-Yau three-folds $X_\pm$. These are shown in figure \ref{fig:TCSG2}. 
Here we summarize the salient points and introduce some notation that will be used in the main text. We refer the reader for a more detailed exposition to  \cite{MR2024648, MR3109862, Corti:2012kd,  Braun:2017uku}. 

The TCS-construction glues these two building blocks times an $S^1$ together, with a hyper-K\"ahler rotation (HKR), or Donaldson matching, $\phi$ acting on the K\"ahler forms $\omega$ and holomorphic (2,0)-forms $\Omega$ as 
\be\label{eq:Dmatching}
\phi:\qquad \left\{ \ba
\omega_\pm &\quad \longleftrightarrow\quad   \Re \Omega^{(2,0)}_{\mp}\cr 
\Im \Omega^{(2,0)}_+ &\quad \longleftrightarrow\quad  -\Im\Omega^{(2,0)} 
\ea\right.
\ee
and an exchange of the circles in the base $\mathbb{S}^1_{e,\pm} = \bbS^1_{b,\mp}$ as shown in figure \ref{fig:TCSG2}. 
In particular $\phi$ induces an isometry on the cohomologies of the fibers: $H^2 (S_+, \Z)\cong H^2 (S_-, \Z)$.

What will be most important for this paper is how to determine the cohomologies of the $G_2$-manifold in terms of data of the TCS gluing. For each building block we define the restriction map on the second cohomology as 
\be
H^{2} (  Z_\pm , \Z)  \quad\stackrel{\rho_\pm}{\longrightarrow}\quad  H^{2} (  S_{\pm}, \Z) \cong \Lambda^{3,19} = U^3 \oplus (-E_8) ^2 \,,
\ee
where we used the cohomology of the K3-surfaces, where $U$ is the hyperbolic lattice of rank 2.
The cohomology of the $G_2$-manifold, can then be written in terms of the following lattices 
\be
\ba
N_\pm &= \hbox{im} (\rho_\pm)\cr 
T_{\pm} & = N_{\pm}^\perp \subset H^2 (S_\pm, \Z)\cr 
K(Z_\pm)  &= \ker(\rho_\pm)/[S_\pm] \,.
\ea \ee
Applying the Mayer-Vietoris sequence to this problem gives the cohomology of the $G_2$ manifold $J$ as
\be\label{TCSMVOUTPUT}
\ba
H^1(J,\Z) & =   0 \cr 
H^2(J,\Z) & =  (N_+ \cap N_-) \ \oplus \ K(Z_+) \ \oplus\  K(Z_-) \cr 
H^3(J,\Z) & = 
\Z[S] \ \oplus\  \Lambda^{3,19} /(N_+ + N_-) \ \oplus\  (N_- \cap T_+)\  \oplus\  (N_+ \cap T_-)\cr 
&\quad  \oplus H^3(Z_+)\ \oplus \ H^3(Z_-) \ \oplus\  K(Z_+) \ \oplus \ K(Z_-) \cr 
H^4(J,\Z) & = 
H^4(S) \oplus 
(T_+ \cap T_-)  \oplus  \Lambda^{3,19} /(N_- + T_+)  \oplus \Lambda^{3,19} /(N_+ + T_-) \cr 
& \quad  \oplus  H^3(Z_+)\oplus H^3(Z_-) \ \oplus \ K(Z_+)^* \oplus K(Z_-)^* \cr 
H^5(J,\Z) & = \Lambda^{3,19} /(T_+ + T_-) \ \oplus K(Z_+)^* \oplus K(Z_-)^* \,.
\ea
\ee
We will determine a similar relation for the $\Spin(7)$ GCS-construction.

\providecommand{\href}[2]{#2}\begingroup\raggedright\endgroup


\end{document}